\documentclass[reprint,amsmath,amssymb,aps,prc,groupedaddress,showpacs,nofootinbib]{revtex4-1}

\usepackage{hyperref}
\usepackage[pdftex]{graphicx}
\usepackage{dcolumn}
\usepackage{bm}
\usepackage{color}

\newcommand{\be}{\begin{equation}}
\newcommand{\ee}{\end{equation}}
\newcommand{\bea}{\begin{eqnarray}}
\newcommand{\eea}{\end{eqnarray}}

\newcommand{\mbss}[1]{_{\mbox{\scriptsize #1}}}

\newcommand{\mbsu}[1]{\mbox{\scriptsize #1}}

\newcommand{\vphu}{\vphantom{*}}
\newcommand{\vphd}{\vphantom{1}}

\newcommand{\ve}{\varepsilon}

\begin{document}

\title{Application of an Extended Random Phase Approximation on Giant Resonances in Light, Medium and Heavy Mass Nuclei}

\author{V. Tselyaev}
\author{N. Lyutorovich}
\affiliation{St. Petersburg State University, 7/9 Universitetskaya nab., St. Petersburg, 198504, Russia}
\author{J. Speth}
\email{J.Speth@fz-juelich.de}
\author{S. Krewald}
\affiliation{Institut f\"ur Kernphysik, Forschungszentrum J\"ulich, D-52425 J\"ulich, Germany}
\author{P.-G. Reinhard}
\affiliation{Institut f\"ur Theoretische Physik II, Universit\"at Erlangen-N\"urnberg,
D-91058 Erlangen, Germany}

\date{\today}

\begin{abstract}
We present results of the time blocking approximation (TBA) on
giant resonances in light, medium and heavy mass nuclei. The TBA
  is an extension of the widely used random-phase approximation (RPA)
 adding complex configurations by coupling to phonon excitations.  A
new method for handling the single-particle continuum is developed and
applied in the present calculations. We investigate in detail the
dependence of the numerical results on the size of the single particle
space and the number of phonons as well as on nuclear matter
properties. Our approach is self-consistent, based on an
energy-density functional of Skyrme type where we used seven different
parameter sets. The numerical results are compared with experimental
data.
\end{abstract}


\maketitle

\section{Introduction} 
Self-consistent mean-field models have developed over the decades
  to a powerful tool for the description of nuclear structure and
  dynamics all over the periodic table
  \cite{Ring1,Bender_2003,Goriely_2002,Nazarevicz1}.  Time-dependent
  mean-field theory allows to simulate a great variety of excitations
  and dynamical processes \cite{Maruhn_2014}.  Giant resonances are
   described well in the small amplitude limit where the space
  of one-particle one-hole $1p1h$ excitations is explored which is, in
  fact, identical to the widely used random phase approximation
  ( RPA). Here one is able to calculate mean energies and total transition strengths.  In
order to describe also the fine structure of bound states and the
total width of giant resonances one has to include correlations beyond
$1p1h$.  Such calculations have been performed in self-consistent as
well as in non-self-consistent approaches.  Extended theories may
include, e.g., two-particle two-hole configurations \cite{Drozdz_1990}
or one may consider the fragmentation of the single-particle states
due to the coupling to phonons
\cite{Krewald_1977,Tselyaev_1989,Tselyaev_2007,Lyutorovich_2015}.
Within the latter approach isoscalar electric monopole resonances and
quadrupole resonances were well reproduced in medium and heavy mass
nuclei
\cite{Kamerdzhiev_1993,Kamerdzhiev_2004,Litvinova_2007,Tselyaev_2007,Lyutorovich_2015}. In
light nuclei like $^{16}$O the present theory is unable to reproduce
the experimental isoscalar cross sections quantitatively as important
decay channels are still missing.  This will be discussed in chapter
III.

One might assume that mean-field theories which describe bulk
properties of nuclei, such as the Thomas-Reiche-Kuhn (TRK) sum rule
and the nuclear symmetry energy \cite{Berman_1975}, as well as shell
effects rather well should also reproduce 
the centroid energies of the giant dipole resonance (GDR).
This is not the case, however, as has worked out in
systematic surveys based on RPA spectra
\cite{Kluepfel_2009,Erler_2011,Erler_2010}. It was impossible to
describe ground-state properties and the centroid energy of the GDR
both in light and heavy nuclei with the same effective
interaction. The problem is more serious than it might appear at a
first glance because the physics of the GDR is closely connected with
the neutron skin thickness and the low-lying dipole strength: the
so-called pygmy resonances
\cite{Abrahamyan_2012,Tamii_2011,Savran_2011}.  These states are
presently investigated experimentally because of their impact on the
isotope abundance produced in supernova explosions
\cite{Horowitz_2001}. 

 Recently we showed that the explicit inclusion  
of quasi particle-phonon coupling may help to solve the problem of mean-field
theories in reproducing the centroid energies of the GDR\cite{Lyutorovich_2012}.  
Within the
time blocking approximation (TBA) \cite{Tselyaev_1989,Tselyaev_2007},
we obtained a reasonably good quantitative agreement with the
experimental data for the GDR in light ($^{16}$O), medium ($^{48}$Ca)
and heavy ($^{208}$Pb) nuclei.  As we went beyond the mean-field
approach we had to adjust new Skyrme forces, where we concentrated on
the GDR in $^{16}$O within the conventional $1p1h$ RPA.  The phonon
contribution did hardly change the $1p1h$ RPA result in $^{16}$O but
moved the GDR in $^{48}$Ca and $^{208}$Pb closer to the experimental
values. The isoscalar giant monopole (GMR) and giant quadrupole
resonances (GQR) were shown in a  short note
\cite{Lyutorovich_2015} using an improved version of TBA that  
derived
all matrix elements consistently from the given (Skyrme)
energy-density functional and calculated them without any
approximations and 
included the single-particle continuum thus avoiding the
artificial discretization implied in earlier TBA calculations.
The present publication discusses in detail the formalism of the  short note \cite{Lyutorovich_2015}. Moreover, we present a new treatment of the single-particle continuum which  allows to include exactly the velocity dependent terms and the 
spin-orbit interaction. We scrutinize the phonon-coupling model by studying the dependence of the results on the numerical parameters of the model (more formal details were presented recently in \cite{Lyutorovich_2016}). The theoretical spectral distributions
for the GMR, GQR and GDR of $^{16}$O, $^{40}$Ca, $^{48}$Ca and
$^{208}$Pb are compared with the experimental ones. We use seven
different Skyrme parametrizations in order to find out how these giant
resonances depend on some specific gross properties of nuclear matter.
As an important result we found that the isoscalar GMR and GQR as well as
isovector GDR can be simultaneously well reproduced by properly chosen Skyrme parametrizations.

The paper is organized as follows. In Chapter II we present in the
Sec. A the basic formulas of the self-consistent RPA and TBA.  In
Sec. B we present seven different Skyrme parametrizations which
reproduce the usual ground-state properties and give reasonably good
results for isovector as well as isoscalar electric giant resonances.
The Skyrme parametrizations were characterized in terms of
  nuclear matter properties (NMP) from which we consider in particular
  four key quantities: incompressibility $K$, effective mass $m^*/m$,
  symmetry energy, and enhancement factor for the TRK sum rule
  $\kappa_\mathrm{TRK}$ (equivalent to isovector effective mass).  We
  investigated in detail the influence of these four NMP on the GDR,
the giant isoscalar monopole and quadrupole resonances. Problems
connected with the tuning of the parameters are discussed in
Sec. C. Details of the calculation scheme are given in Chapter III.
In Sec. A we discuss the single-particle basis and in Sec. B the
effect of the exact continuum treatment on our results.  In Sec. C we
investigate in detail the dependence of the TBA results on the number
of phonons included. Chapter IV presents our results. In Sec. A the
impact of the phonon coupling on the resonances is shown and in Sec. B
we compare our final results with experimental data. In the last
chapter we summarize our investigations.

\section{The method}
\label{sect2}
\subsection{The basic equations}
\label{sect2a} 

\subsubsection{Conventional RPA}

The original derivation of the RPA equations in nuclear physics is
based on the time-dependent Hartree-Fock methods where one considered
small amplitude dynamics about a Hartree-Fock ground state \cite{Bro71aB}.  From this derivation, one may obtain the impression that the RPA is a very limited approach.  This is actually not the case if one considers the derivation within the Green function
method.  All details and the explicit expressions can be found in Ref. \cite{Speth_1977}.
The transition matrix element of a one-particle operator 
between the exact ground state of an $A$-particle system 
and an excited state $m$ is given as:
\begin{equation}\label{eq:1a}
\langle Am|Q|A0 \rangle \;=\sum_{\nu_1 \nu_2}Q^{\rm eff}_{\nu_1 \nu_2}\; 
\chi ^{m}_{\nu_1 \nu_2}.
\end{equation} 
Here Q$^{\rm eff}$ are effective operators and $\chi^m$ are 
the quasiparticle-quasihole matrix elements which are given by the equation:
\bea\label{eq:1b}
\left( \epsilon_{\nu_1}-\epsilon_{\nu_2}- \Omega\right)\chi^{m}_{\nu_1 \nu_2} =
\nonumber \\
\left(n_{\nu_1}-n_{\nu_2}\right)\sum_{\nu_3 \nu_4}
F^{ph}_{\nu_1 \nu_4 \nu_2 \nu_3} \;\chi^{m}_{\nu_3 \nu_4}
\eea
where F$^{ph}$ is the renormalized $ph$ interaction. 
All relations have been been derived without any approximations. 
Therefore conservation laws can be applied. 
E.g., the effective electric operators reduces to the bare ones 
due to Ward identities in the long-wave length limit.
The derivation of the RPA equation starts with the equation of motion 
(Dyson equation) for the one-particle Green function. 
The basic input is the mass operator $\Sigma$ which include 
all information on the many-body system. The most general form is given as:
\bea \label{1c}
\Sigma = \Sigma {(\mathbf{r} ,\mathbf{p},\epsilon)} 
\eea
It depends on the coordinate $\mathbf{r}$, the momentum $\mathbf{p}$ 
(non-locality), and the energy $\epsilon$.

Note: the RPA equations derived here are formally identical with the
corresponding equations derived in the linear response limit of
  time-dependent density-functional theory (TDDFT) in the next
section. The crucial difference is the mass operator in Eq. \ref{1c}
which is energy dependent in a general many-body theory whereas
  it turns out to be independent of energy in (TDDFT). As the various
quantities in the general case and in linear response are different,
we also use different symbols.

In the general case, the expression for the effective mass has the
form:
\begin{equation}\label{eq:50}
\frac{m}{m^*} =
\frac{\left(1+2m \frac{\delta \Sigma}{\delta p^2}\right)_{F}} 
{\left({1-\frac{\delta \Sigma}{\delta \epsilon}}\right)_F}.
\end{equation}
The nominator is called $k$-mass and the denominator $E$-mass
\cite{Jeukenne_1976}.  They are related to the non-locality and
energy-dependence of the mass operator, respectively.  If the mass
operator does not depend on the energy, the denominator is equal to
one.  In the case of a totally energy independent mass operator,
the formulas become much simpler as the single-particle strength is
equal to one \cite{Frank_2006}. The effective operators are in all
cases equal to the bare operators and also the $ph$-interaction is not
renormalized.

In our extended model (the TBA), we introduce complex
  configurations by coupling phonons to the single-particle states.
This introduces an energy dependence into the mass operator in first
order \cite{Wambach_1982}.  For this reason the single-particle 
strength is less then one and we obtain a contribution to the
$E$-mass.  This is the well known shift due to phonon coupling.
All this is correctly taken care of in the TBA. But we will not
address single-particle effects explicitely later on.

\subsubsection{Self-consistent RPA}

Our approach is based on the version of the response function formalism
developed within the Green function method (see Ref.~\cite{Speth_1977}).
In the general case the distribution of the strength of transitions
in the nucleus caused by some external field represented by the
single-particle operator $Q$ is determined by the strength function
$S(E)$ which is defined in terms of the response function $R(\omega)$
by the formulas
\be
S(E)=-\frac{1}{\pi}\;\mbox{Im}\,\Pi(E+i\Delta)\,,
\label{sfdef}
\ee
\be
\Pi(\omega) = - \langle\,Q\,|\,R(\omega)\,|\,Q\,\rangle\,,
\label{poldef}
\ee
where $E$ is an excitation energy, $\Delta$ is a smearing parameter,
and $\Pi(\omega)$ is the (dynamic) polarizability.

The first model used in our calculations is the self-consistent RPA
based on TDDFT with the energy density
functional $E[\rho]$. The TDDFT equations imply that
$[\,\rho,h\,]=0$ where $\rho$ is the single-particle density matrix
satisfying the condition $\rho^2=\rho$, and $h$ is the single-particle
Hamiltonian,
\be
h^{\vphu}_{12} = \frac{\delta E[\rho]}{\delta\rho^{\vphu}_{21}}\,.
\label{sphdft}
\ee
The numerical indices here and in the following denote the set of the
quantum numbers of some single-particle basis.  It is convenient to
introduce the basis that diagonalizes simultaneously the
operators $h$ and $\rho\,$:
\be
h^{\vphu}_{12} = \ve^{\vphu}_{1}\delta^{\vphu}_{12}\,,
\qquad
\rho^{\vphu}_{12} = n^{\vphu}_{1}\delta^{\vphu}_{12}\,,
\label{spbas}
\ee
where $n^{\vphu}_{1}=0,1$ is the occupation number.
In what follows the indices $p$ and $h$ will be used to label
the single-particle states of the particles ($n^{\vphu}_{p} = 0$)
and holes ($n^{\vphu}_{h} = 1$) in this basis.

In RPA, the response function is a solution of the following
Bethe-Salpeter equation (BSE)
\be
R^{\mbss{RPA}}_{\vphd}(\omega) = R^{(0)}_{\vphd}(\omega)
- R^{(0)}_{\vphd}(\omega)VR^{\mbss{RPA}}_{\vphd}(\omega)\,,
\label{rfrpa}
\ee
where $R^{(0)}_{\vphd}(\omega)$ is the uncorrelated $1p1h$
propagator and $V$ is the residual interaction.
The $1p1h$ propagator $R^{(0)}_{\vphd}(\omega)$ is defined as
\be
R^{(0)}_{\vphd}(\omega) = -
\bigl(\,\omega - \Omega^{(0)}_{\vphd}\bigr)^{-1}
M^{\mbss{RPA}}_{\vphd},
\label{rf0}
\ee
where the matrices $\Omega^{(0)}_{\vphd}$ and $M^{\mbss{RPA}}_{\vphd}$
are defined in the $1p1h$ configuration
space. $M^{\mbss{RPA}}_{\vphd}$ is the metric matrix
\be
M^{\mbss{RPA}}_{12,34} =
\delta^{\vphu}_{13}\,\rho^{\vphu}_{42} -
\rho^{\vphu}_{13}\,\delta^{\vphu}_{42}\,.
\label{mrpa}
\ee
The matrix $\Omega^{(0)}_{\vphd}$ has the form
\be
\Omega^{(0)}_{12,34} =
h^{\vphu}_{13}\,\delta^{\vphu}_{42} -
\delta^{\vphu}_{13}\,h^{\vphu}_{42}\,.
\label{omrpa}
\ee
In the self-consistent RPA based on the energy density functional
$E[\rho]$ one has
\be
{V}^{\vphu}_{12,34} =
\frac{\delta^2 E[\rho]}
{\delta\rho^{\vphu}_{21}\,\delta\rho^{\vphu}_{34}}\,,
\label{sccond}
\ee
so the quantities $h$ and $V$ appear to be linked by Eqs. (\ref{sphdft})
and (\ref{sccond}).

The propagator $R^{\mbss{RPA}}_{\vphd}(\omega)$, being a matrix in $1p1h$ space, is a rather bulky object. For practical calculations, it
is more convenient to express it in terms of RPA amplitudes $z_{12}^n$
by virtue of the spectral representation
\be
  R^{\mbss{RPA}}_{1234}(\omega)
  =
  -\sum_nz_{12}^n\frac{\mbox{sgn}(\omega_n)}{\omega-\omega_n}(z_{34}^n)^*
\ee
where $n$ labels the RPA eigenmodes and $\omega_n$ is the eigenfrequency.
Inserting that into Eq.~(\ref{rfrpa}) and filtering the pole at
$\omega=\omega_n$ yields the familiar RPA equations 
\be
 \sum_{34}\left(\Omega^{(0)}_{12,34}
    + \sum_{56} M^{\mbss{RPA}}_{12,56}\,{V}^{\vphu}_{56,34}\right)
    \,z^{n}_{34} 
=
  \omega^{\vphu}_n\,z^{n}_{12}\,,
\label{rpaze}
\ee
where the transition amplitudes ${z}^{n}$ are normalized
by the condition
\be
 \sum_{12,34}({z}^{n}_{12})^*\,M^{\mbss{RPA}}_{1234}\,{z}^{n'}_{34} 
 =
 \mbox{sgn}(\omega^{\vphu}_{n})\,\delta^{\vphu}_{n,\,n'}.
\label{zmz}
\ee
These equations determine the set of eigenstates $n$ with amplitudes
$z^{n}_{12}$ and frequencies $\omega_n$.

\subsubsection{Phonon coupling model}

The second model is the quasiparticle-phonon coupling model within the
time-blocking approximation (TBA) \cite{Tselyaev_1989,
  Kamerdzhiev_1997, Kamerdzhiev_2004, Tselyaev_2007} (without ground
state correlations beyond the RPA included in \cite{Tselyaev_1989,
  Kamerdzhiev_1997, Kamerdzhiev_2004, Tselyaev_2007} and without
pairing correlations included in \cite{Tselyaev_2007}).  This model,
which in the following will be referred to as TBA, is an extension of
RPA including $1p1h\otimes$phonon configurations in addition to the
$1p1h$ configurations incorporated in the conventional RPA.  The BSE
for the response function in the TBA is
\begin{eqnarray}
  R^{\mbss{TBA}}_{\vphd}(\omega) 
  &=&
  R^{(0)}_{\vphd}(\omega) \nonumber \\
  &&- 
  R^{(0)}_{\vphd}(\omega)(V\!+\!\tilde{W}(\omega))
  R^{\mbss{TBA}}_{\vphd}(\omega) \,,
\label{rftba}\\
  \tilde{W}(\omega) 
  &=& W(\omega)-{W}(0)\,,
\label{Wsubtract}
\end{eqnarray}
where the induced interaction $\tilde{W}(\omega)$ 
serves to include contributions of $1p1h\otimes$phonon configurations.

The matrix ${W}(\omega)$ in Eq. (\ref{Wsubtract}) is defined in the
$1p1h$ subspace and can be represented in the form
\be
{W}^{\vphu}_{12,34}(\omega) =
\sum_{c,\;\sigma}\,\frac{\sigma\,
{F}^{c(\sigma)}_{12}
{F}^{c(\sigma)*}_{34}}
{\omega - \sigma\,\Omega^{\vphu}_{c}}\,,
\label{wdef}
\ee
where $\sigma = \pm 1$, $\,c = \{p',h',n\}$ is an index of the subspace
of $1p1h\otimes$phonon configurations, $n$ is the phonon's index,
\be
\Omega^{\vphu}_{c} =
\ve^{\vphu}_{p'} - \ve^{\vphu}_{h'} + \omega^{\vphu}_{n}\,,
\quad \omega^{\vphu}_{n}>0\,,
\label{omcdef}
\ee
\be
{F}^{c(-)}_{12}={F}^{c(+)*}_{21},\qquad
{F}^{c(-)}_{ph}={F}^{c(+)}_{hp}=0\,,
\label{fcrel}
\ee
\be
{F}^{c(+)}_{ph} =
\delta^{\vphu}_{pp'}\,g^{n}_{h'h} -
\delta^{\vphu}_{h'h}\,g^{n}_{pp'},
\label{fcdef}
\ee
$g^{n}_{12}$ is an amplitude of the quasiparticle-phonon interaction.
These $g$ amplitudes (along with the phonon's energies
$\omega^{\vphu}_{n}$) are determined by the positive frequency
solutions of the RPA equations and the emerging $z$ amplitudes as
\be
g^{n}_{12} = \sum_{34} {V}^{\vphu}_{12,34}\,z^{n}_{34}\,.
\label{gndef}
\ee
where $V_{12,34}$ is the same residual interaction (13) as used in RPA.
In our DFT-based approach the energy density functional $E[\rho]$ in
Eqs. (\ref{sphdft}) and (\ref{sccond}) is the functional of the Skyrme
type with free parameters which are adjusted to experimental data. In
this case $E[\rho]$ already effectively contains a part (actually
  the stationary part) of the contributions of those
$1p1h\otimes$phonon configurations which are explicitly included in
the TBA. Therefore, in the theory going beyond the RPA, the problem of
double counting and of ground-state stability arises
\cite{Toe88a}.  To avoid this problem in the TBA, we use the
subtraction method.  It consists in the replacement of the amplitude
${W}(\omega)$ by the quantity $\bar{W}(\omega)={W}(\omega)-{W}(0)$ as
it is given in Eq.~(\ref{rftba}).  In Ref.~\cite{Tselyaev_2013} it was
shown that, in addition to the elimination of double counting, the
subtraction method ensures stability of solutions of the TBA
eigenvalue equations.

\subsection{Basics on the Skyrme functional and related parameters}
\label{sec:Skyrme}

From the variety of self-consistent nuclear mean-field models
\cite{Bender_2003}, we consider here a non-relativistic branch, the
widely used and very successful Skyrme-Hartree-Fock (SHF) functional.
A detailed description of the functional is found in the reviews
\cite{Bender_2003,Stone_2007,Erler_2011}. We summarize the essential
features: The functional depends on a couple of local densities and
currents (density, gradient of density, kinetic-energy density,
spin-orbit density, current, spin density, kinetic spin-density). It
consists of quadratic combinations of these local quantities,
corresponding to pairwise contact interactions. The term with the
local densities is augmented by a non-quadratic density dependence to
provide appropriate saturation. One adds a simple pairing functional
to account for nuclear superfluidity.  The typically 13--14 model
parameters are determined by a fit to a large body of experimental
data on bulk properties of the nuclear ground state. For recent
examples see \cite{Goriely_2002,Kluepfel_2009,Kortelainen_2010}.

The properties of the forces can be characterized, to a large extend, by
nuclear matter properties (NMP), i.e. equilibrium properties of
homogeneous, symmetric nuclear matter, for which we have some
intuition from the liquid-drop model \cite{Myers_1977}.  Of
particular interest for resonance excitations are the NMP which are
related to response to perturbations: incompressibility $K$ (isoscalar
static), effective mass $m^*/m$ (isoscalar dynamic), symmetry energy
$a_\mathrm{sym}$ (isovector static), TRK sum rule enhancement
$\kappa_\mathrm{TRK}$ (isovector dynamic). We aim at exploring the
effect of phonon coupling under varying conditions and thus use here
parametrizations from recent fits presented in \cite{Kluepfel_2009}
which provides a systematic variation of these four NMP.

\begin{table}
\centering
\begin{tabular}{l|cccc}
  & $K$ [MeV]& $m^*/m$ & $a_\mathrm{sym}$ [MeV]& $\kappa_\mathrm{TRK}$ \\
\hline
SV-bas &  234 &  0.90  &  30  &  0.4 \\
SV-kap00 & 234 &  0.90  &  30  &  0.0 \\
SV-mas07 & 234 &  0.70  &  30  &  0.4 \\
SV-sym34 & 234 &  0.90  &  34  &  0.4 \\
SV-K218 & 218 &  0.90  &  30  &  0.4 \\
SV-m64k6 & 241 & 0.64  & 27 & 0.6\\
SV-m56k6 &  255 &   0.56 &   27 &   0.6\\
\hline
\end{tabular}
\caption{\label{tab:NMP}
Nuclear matter properties for the Skyrme paramterizations
used in this study: incompressibility $K$, isoscalar effective mass
$m^*/m$, symmetry energy $a_\mathrm{sym}$, Thomas-Reiche-Kuhn sum rule
enhancement $\kappa_\mathrm{TRK}$. The first five
parametrizations stem from  \cite{Kluepfel_2009}, the last two from
\cite{Lyutorovich_2012}.}

\end{table}
Table \ref{tab:NMP} lists the selection of parametrizations and their
NMP. SV-bas is the base point of the variation of forces. Its NMP are
chosen such that dipole polarizability and the three most important
giant resonances (GMR, GDR, and GQR) in $^{208}$Pb are well reproduced
by Skyrme-RPA calculations. Each one of the next four parametrizations
vary exactly one NMP while keeping the other three at the SV-bas
value. These 1+4 parametrizations allow to explore the effect of each
NMP separately.  It was figured out in \cite{Kluepfel_2009} that there
is a strong relation between each one of the four NMP and one specific
giant resonance: $K$ affects mainly the GMR, $m^*/m$ affects mainly
the GQR, $\kappa_\mathrm{TRK}$ affects the GDR, and $a_\mathrm{sym}$
is linked to the dipole polarizability \cite{Nazarewicz_2013}.

Finally, the last two parametrizations in Table \ref{tab:NMP} were
developed in \cite{Lyutorovich_2012} with the goal to describe, within
TBA, at the same time the GDR in $^{16}$O and $^{208}$Pb. This
required to push up the RPA peak energy which was achieved by low
$a_\mathrm{sym}$ in combination with high $\kappa_\mathrm{TRK}$. To
avoid unphysical spectral distributions for the GDR, a very low
$m^*/m$ was used.

\subsection{The problem of tuning a parametrization}
\label{sec:tuning}

Looking only on average resonance energies, the tuning of
parametrizations is simple. As mentioned before, the three giant resonances which we
consider couple each one almost exclusively to one property, the GMR
to the incompressibility $K$, the GDR to the TRK sum rule enhancement
$\kappa_\mathrm{TRK}$, and the GQR to the isoscalar effective mass
$m^*/m$. This suggests that one can adjust these three resonances
independently at wish. However, problems appear when looking at the
detailed spectral distributions. We observed in our investigations
that the shift in average resonance energies does usually not
correspond to a global shift of the spectral distribution, but rather
to a redistribution of strength over the spectrum.  However, such
redistribution can lead to unrealistic profiles and that is what is
often hindering a light-hearted adjustment.

\begin{figure}
\centerline{\includegraphics[width=0.99\linewidth]{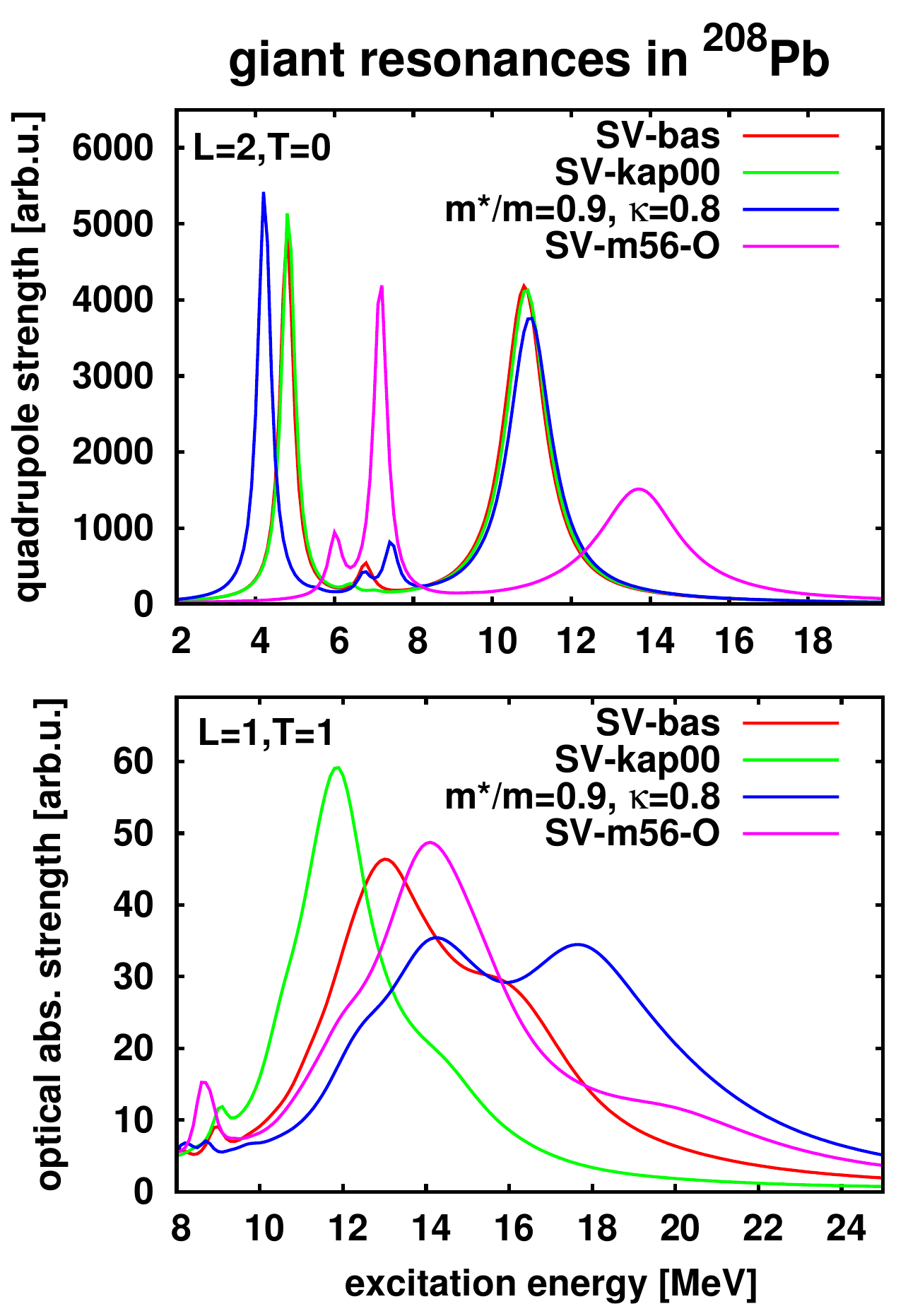}}
\caption{\label{fig:GR} Dipole strength (lower panel) and quadrupole
  strength (upper panel) for four parametrizations as indicated. The
  smooth spectra are obtained from folding with Gaussians of linearly
  increasing width $\Gamma=\mbox{max}(0.2,(E-8)/5)$ MeV. }
\end{figure}
Figure \ref{fig:GR} shows detailed spectra for four parametrizations.
SV-kap00 as compared to SV-bas corresponds to a shift of
$\kappa_\mathrm{TRK}$ from 0.4 (for SV-bas) down to 0. This has no
effect on the GQR and leads to a visible downshift of the GDR. This
downshift does also change the profile to the extend that high-energy
bump at 16 MeV in SV-bas now appears at 14 MeV and, more important,
becomes much smaller. Thus the way from SV-kap00 to SV-bas already
changes somewhat the profile, but at a harmless level.

Now we try to up-shift the GDR by enhancing dramatically
$\kappa_\mathrm{TRK}$ to 0.8 while keeping $m^*/m=0.9$ at the value of
SV-bas. This leads to the blue curves in the figure. It is gratifying
to see that the GQR remains where it should be. The GDR makes the
wanted up-shift. However, this happens at the price of a totally
unrealistic double humped structure of the GDR. Mind that the upper
bump appears in so pronounced manner in spite of energy-dependent
folding width. Mere enhancement of $\kappa_\mathrm{TRK}$ seems thus no
solution to the wanted up-shift of the GDR. The former solution was to
use much lower $m^*/m=0.56$ to curb down the double hump. This is
successful for the GDR (purple line) however disastrous for the
GQR. Not only that the too high GQR position cannot be cured by phonon
coupling, but also that the low energy spectrum is grossly
unrealistic. This looks like a deadlock for global improvment
  and it is at the level of RPA. The situation becomes more gracefull
  for TBA as we will see later.

\section{Details of the calculation scheme}
\label{sec:details of the calculation}

\subsection{Single-particle basis and residual interaction}
\label{sec:calc}
  The response functions both for RPA and TBA, Eqs. (\ref{rfrpa}) and (\ref{rftba}),   are solved in a discrete basis defined as a
set of solutions of the Schr\"odinger equation with box boundary
conditions. Both equations are solved in the same large configuration
space. A new method to include  the continuum in the discrete basis representation
is explained in Appendix A. The residual interaction $V$ in
Eqs.(\ref{rfrpa}) and (\ref{rftba}) is derived from the energy
functionals according to Eq.~(\ref{sccond}).  In the case of the
energy density functional $E[\rho]$ built on the Skyrme forces, the
amplitude $V$ determined by Eq.~(\ref{sccond}) contains zero-range
(velocity-independent) and velocity-dependent parts.  The scheme for
taking into account the zero-range part of the residual interaction
adopted in our calculations is described in Refs.~\cite{Litvinova_2007,Tselyaev_2007}. A detailed description of the computation of the matrix elements in connection with 
the Skyrme functional is found in \cite{Lyutorovich_2016}.

We will consider only doubly-magic nuclei. They have closed shells and
pairing is inactive. The box sizes in the RPA and TBA calculations are
15 fm for $^{16}$O, $^{40,48}$Ca and 18 fm for $^{208}$ Pb. The
single-particle basis in which we solve the RPA and TBA equations
include single-particle states up to $\varepsilon_{\text{max}}$
= 100 MeV (see our discussion in the next two sections).  In the TBA
calculations we apply the subtraction recipe (\ref{Wsubtract})
\cite{Tselyaev_2013}.  As mentioned before, this procedure eliminates
double counting, resolves stability problems, and restores the
Thouless theorem.

\subsection{Effect of the exact Continuum}

\begin{figure}
\centerline{\includegraphics[width=10cm]{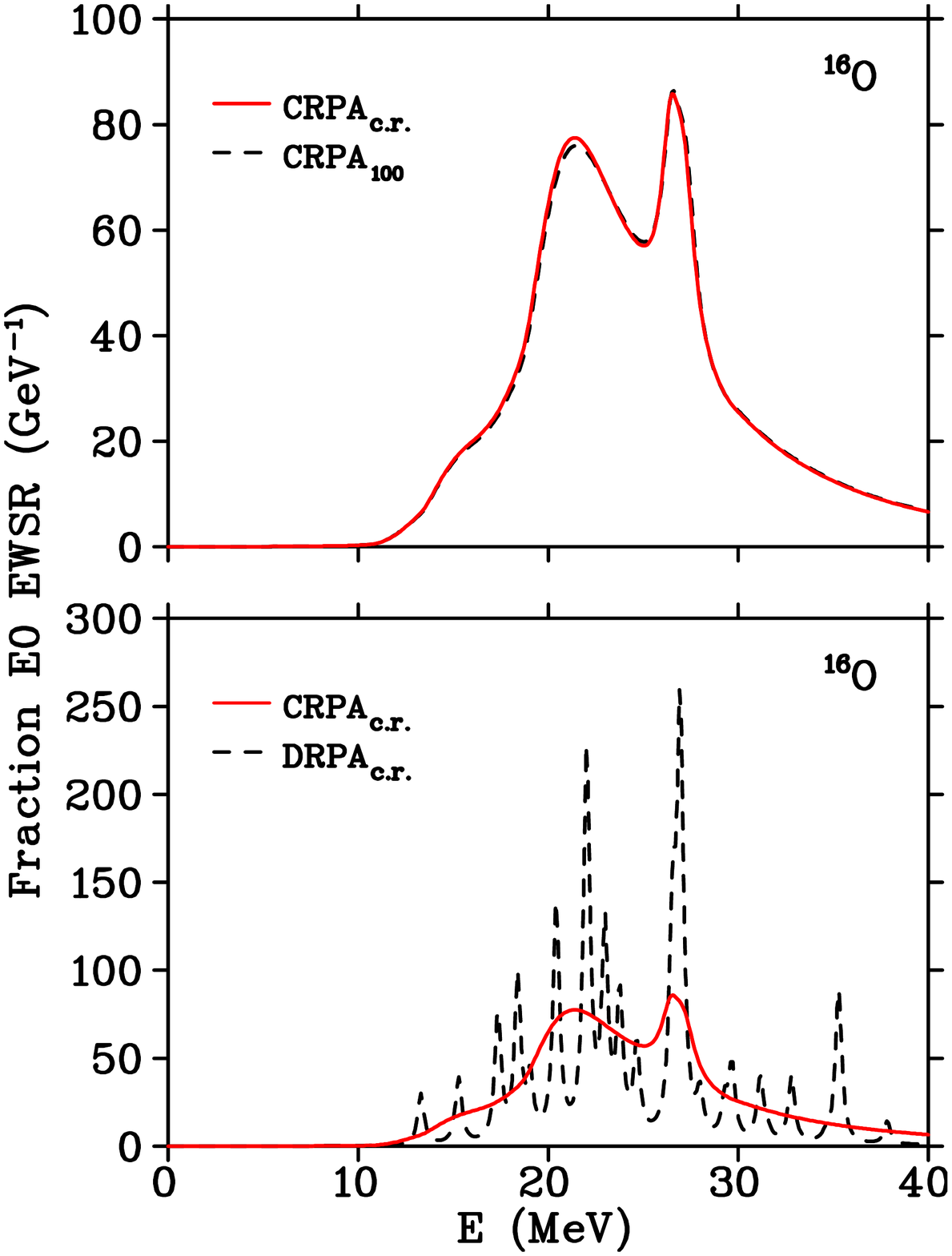}}
\caption{\label{fig:o16e0rpa}
ISGMR in $^{16}$O calculated within fully self-consistent RPA based
on the Skyrme energy density functional with the T6 parametrization
\cite{TBFP}. Fractions of the $E0$ EWSR are shown.
Upper panel:
the CRPA$_{\,\mbsu{c.r.}}$ function obtained by making use of the method
of Ref.~\cite{Tselyaev_2009} is presented by the red solid line.
The CRPA$_{\,\mbsu{100}}$ function obtained in the discrete basis
with $\varepsilon_{\text{max}}$ = 100~MeV is presented by the black dashed line.
Lower panel:
the CRPA$_{\,\mbsu{c.r.}}$ function (red solid line) is compared with
the DRPA$_{\,\mbsu{c.r.}}$ function (black dashed line) obtained
by the same method of Ref.~\cite{Tselyaev_2009}.
Smearing parameter $\Delta$ = 200~keV was used in all the calculations.}
\end{figure}

As mentioned before, we included the full single-particle continuum
into our TBA calculations. For this, we use a new technique which allows a continuum treatment in connection with full self-consistency RPA as outlined in Appendix \ref{app:cont}.  This method uses the discrete basis representation and recovers the exact
method \cite{Shlomo_1975} of treatment of the continuum in the
coordinate representation if the discrete basis is sufficiently
complete ($\varepsilon_{\text{max}}$ high enough) and the
radius of the box ($R_{\mbsu{box}}$) is sufficiently large (see
Appendix \ref{app:cont}).
To check the accuracy of our method, we first compare the results
obtained within the continuum RPA (CRPA) in the discrete basis
representation (hereafter called CRPA$_{\,\mbsu{d.b.}}$) with the
results of the CRPA in the coordinate representation
(CRPA$_{\,\mbsu{c.r.}}$).  As an example, we consider calculations of
the GMR in the fully self-consistent CRPA based on the Skyrme energy
density functional with the T6 parametrization \cite{TBFP} producing
the nucleon effective mass $m^*/m=1$. As was shown in
Ref.~\cite{Tselyaev_2009}, the fully self-consistent
CRPA$_{\,\mbsu{c.r.}}$ scheme in this special case has relatively
simple form.  The results for the nucleus $^{16}$O are shown in
Fig.~\ref{fig:o16e0rpa}.  The function $F(E)$ presented in this figure
is the fraction of the energy-weighted sum rule (EWSR) defined as
\be
F(E)=E\,S(E)/m_{1}\,,
\label{fredef}
\ee
where $S(E)$ is the strength function defined in Eq.(\ref{sfdef}) and
$m_{1}=\int dE,\,E\,S(E)$ is the energy-weighted moment of $S(E)$
determined by the known EWSR \cite{BLM79}.

In the upper panel of Fig. \ref{fig:o16e0rpa} the
CRPA$_{\,\mbsu{c.r.}}$ results are compared with CRPA$_{\,\mbsu{100}}$
obtained in the discrete basis with $E_{\mbsu{cut}}$ = 100~MeV.  The
equations of the CRPA$_{\,\mbsu{c.r.}}$ were solved with a mesh
  spacing $h=0.05$~fm in $r$-space and box size
  $R_{\mbsu{box}}=15$~fm. All these calculations used a smearing
parameter $\Delta$ = 200 keV. 
The difference between the CRPA$_{100}$ and CRPA$_{\,\mbsu{c.r.}}$ curves is small and hardly visible.The CRPA$_{300}$ obtained in the discrete basis with $E_{\mbsu{cut}}$
= 300~MeV and CRPA$_{\,\mbsu{c.r.}}$ curves are practically
indistinguishable, so we do not show them.  In the lower panel of
Fig. \ref{fig:o16e0rpa} the discrete RPA (DRPA) results obtained by the
coordinate representation method of Ref.~\cite{Tselyaev_2009} are
compared with the CRPA function for $^{16}$O and, again, $\Delta$ =
200~keV.  In this case, the difference between these results is large.

\begin{figure*}
\centerline{\includegraphics[width=0.9\linewidth,angle=90]{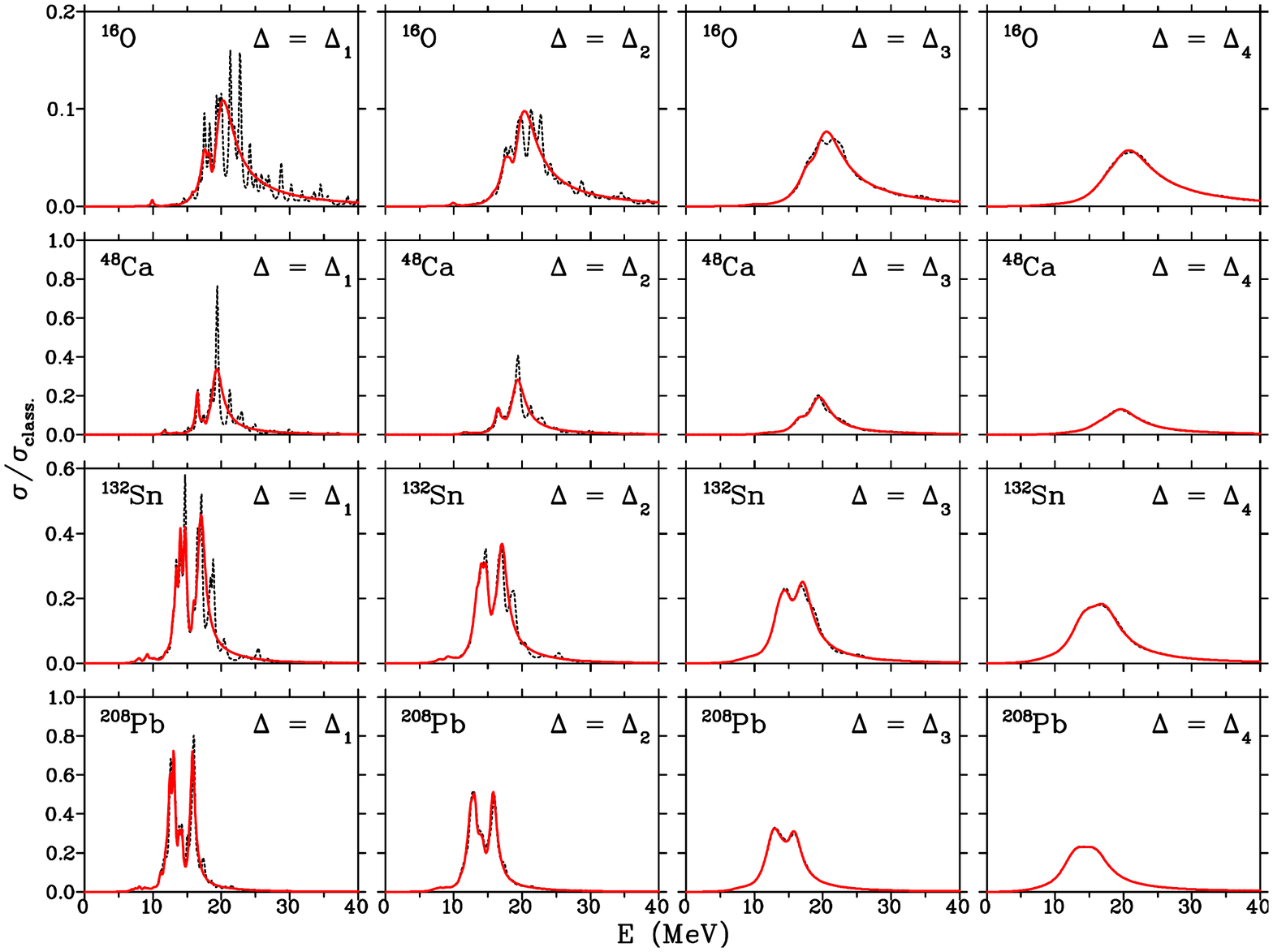}}
\caption{\label{figcvd} (Color online.)
Photo-absorption cross sections in the nuclei $^{16}$O, $^{48}$Ca, $^{132}$Sn,
and $^{208}$Pb calculated within the CRPA (red solid lines) and the DRPA
(black dashed lines) with different smearing parameters $\Delta$:
$\Delta_1 = 200$ keV, $\Delta_2 = 400$ keV, $\Delta_3 = 1$ MeV, and
$\Delta_4 = 2$ MeV.
The discrete basis representation with $E_{\mbsu{cut}} = 100$ MeV is
used both in the CRPA and the DRPA.
The calculated cross sections have been normalized to the classical
values $\sigma_{\mbsu{class.}} = \frac{5}{3}\pi\langle r^2 \rangle$
(see text for more details).
The results are obtained with the SV-bas Skyrme force parametrization
\cite{Kluepfel_2008}.
}
\end{figure*}

Thus we see that the magnitude of the continuum effects on
nuclear excitations is different in light and heavy nuclei. To see the trend we have calculated the GDR in the nuclei $^{16}$O,
$^{48}$Ca, $^{132}$Sn and $^{208}$Pb within two schemes:
CRPA$_{\,\mbsu{d.b.}}$ and DRPA$_{\,\mbsu{d.b.}}$ using the
  Skyrme parametrization SV-bas \cite{Kluepfel_2008}.  The results
are presented in Fig.~\ref{figcvd}.  In this figure, the
photo-absorption cross sections normalized to the classical values
$\sigma_{\mbsu{class.}} = \frac{5}{3}\pi\langle r^2 \rangle$ are
shown. The mean-square radii $\langle r^2 \rangle$ have been
calculated for the each nucleus using its Skyrme-Hartree-Fock
ground-state.  The $\sigma_{\mbsu{class.}}$ are: 378.5~mb for
$^{16}$O, 654.5~mb for $^{48}$Ca, 1204.0~mb for $^{132}$Sn, and
1605.1~mb for $^{208}$Pb.  As can be seen, the effect of the
single-particle continuum is strongest in the light nuclei $^{16}$O
and $^{40}$Ca. In $^{16}$O nucleus, the CRPA and DRPA results
significantly differ at $\Delta\lesssim$ 400~keV. Even at $\Delta =$
1~MeV the difference is noticeable. It disappears only at $\Delta =$
2~MeV.  In $^{48}$Ca the difference between the CRPA and DRPA becomes
small at $\Delta\gtrsim$ 1~MeV.  The same is true for $^{132}$Sn,
though in whole this difference here is less than in $^{48}$Ca.  These
results are in agreement with the conclusions of
Refs. \cite{Krewald_1977,Nakatsukasa_2005,DeDonno_2011}.  In the heavy
nucleus $^{208}$Pb, the effect of the single-particle continuum is
small and is manifested only at $\Delta\lesssim$ 200~keV.

\begin{figure}
\centerline{\includegraphics[width=9cm]{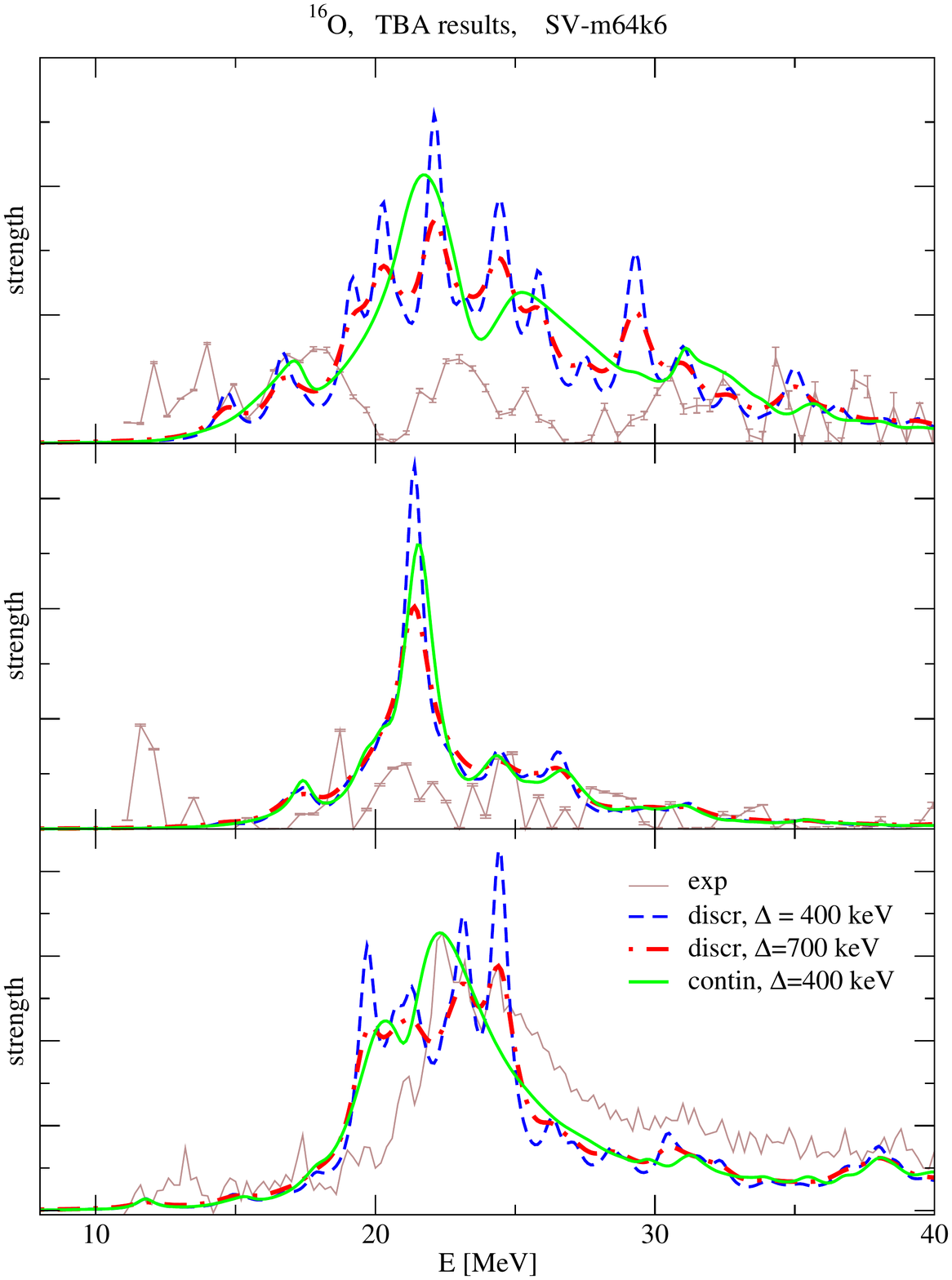}}
\caption{\label{fig:16O_dtba_ctba_m64k6}
Discrete and continuum TBA results for $^{16}$O which were obtained
with the parameter set SV-m64k6. The fractions EWSR for GMR and GQR
and photo-absorption cross section for GDR are presented in the upper,
middle and lower panel, respectively.
The DTBA for smearing parameters $\Delta$ = 400 and 700 keV
are given by blue dashed and red dashed-doted lines, respectively.
Thick green and thin brown full lines represent CTBA with $\Delta$ = 400 keV
and experimental data, respectively. The data are taken from Ref. \cite{Ishkhanov_2002,Lui_2001}}
\end{figure}

\begin{figure}
\centerline{\includegraphics[width=9cm]{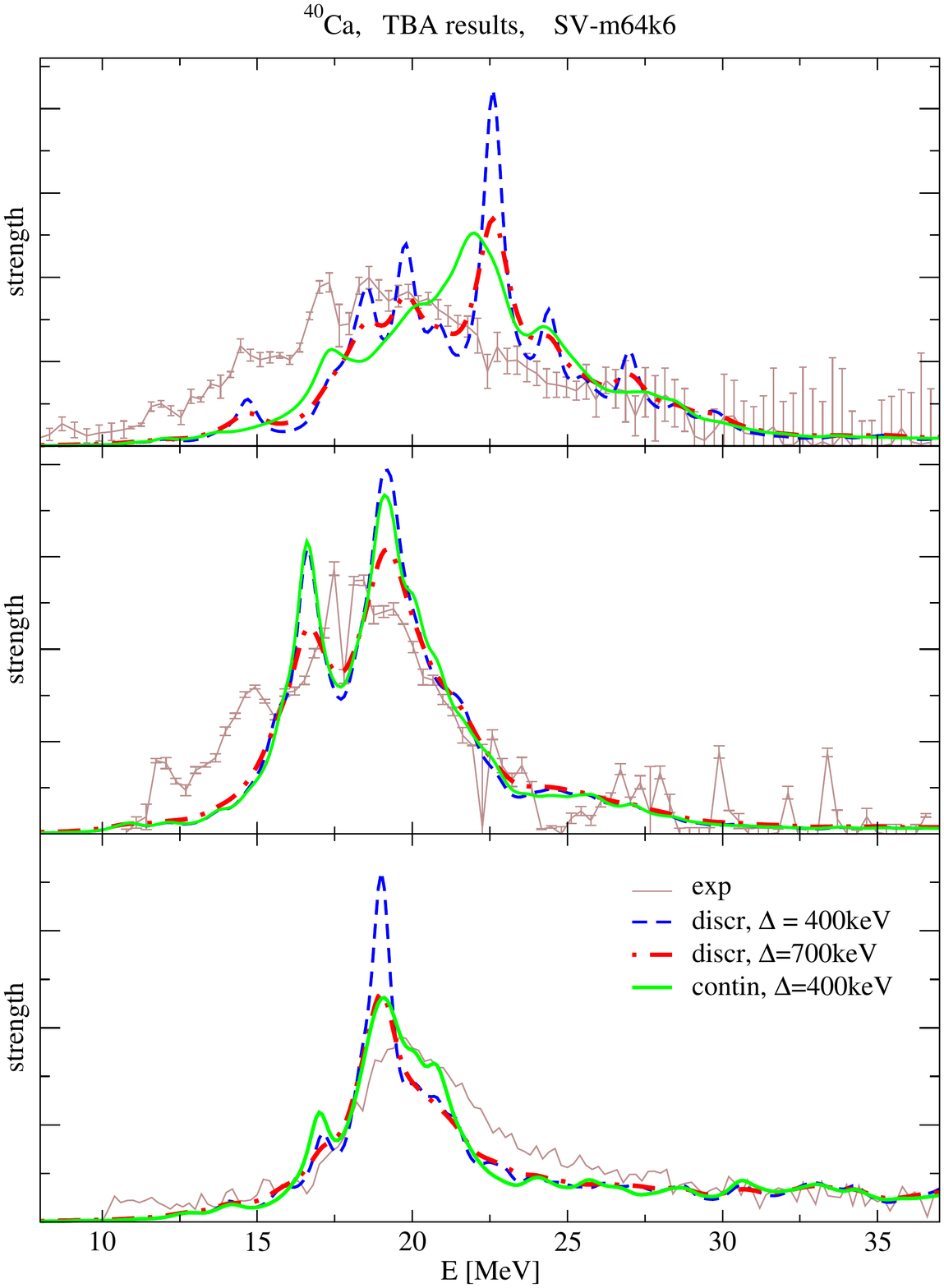}}
\caption{\label{fig:40Ca_dtba_ctba_m64k6}
Same as in Fig.~\ref{fig:16O_dtba_ctba_m64k6} but for $^{40}$Ca. The corresponding data are taken from Ref. \cite{Erokhova_2003,Anders_2013}}
\end{figure}

In Fig.~\ref{fig:16O_dtba_ctba_m64k6}, for $^{16}$O, and
Fig.~\ref{fig:40Ca_dtba_ctba_m64k6}, for $^{40}$Ca, we compare the TBA
results obtained with the exact continuum treatment (CTBA) and the
discretized approximation (DTBA). Here, blue dashed and red
dashed-dotted lines represent the DTBA results for smearing parameters
$\Delta$ = 400 and 700 keV, respectively. The expression "strength" in
the Y-axes mean fractions EWSR for GMR and GQR and photo-absorption
cross section for GDR.  The experimental data for GMR and GQR in
$^{16}$O were taken from Ref.~\cite{Lui_2001} and for GDR in $^{16}$O
from \cite{Ishkhanov_2002}. The data for $^{40}$Ca were taken from
Refs.~\cite{Anders_2013} and \cite{Erokhova_2003}, respectively.  The
figures show that, for light nuclei, increasing $\Delta(\mbox{DTBA})$
damps the artificial fine structure of the discrete approach.  But, at
the same time, it wipes out important physical features. Hence, it is
impossible to reproduce CTBA results for strength functions of light
nuclei by using the DTBA, both with small and large smearing
parameters.

The experimental profiles for the two isoscalar resonances in $^{16}$O
look very different from the isovector GDR and from all resonances in
heavier nuclei. The theoretical GQR shows a narrow peak where as the
experimental strength is nearly continuously distributed over more
then 20 MeV. The same is true also for the experimental GMR strength.
Here the theoretical strength distribution is very broad and shows at
least some qualitative similarity.  There are little differences
between the various parameter sets.  The question arises why are we
not able to reproduce theoretically these two resonances while the
results in the heavier nuclei are in good qualitative in many cases
even in quantitative agreement with experiment? For the GQR the
explanation is simple: The dominant decay channel of the GQR in
$^{16}$O is the $\alpha$-decay into the ground state and the first
excited state of $^{12}$C \cite{Wagner}.
In the range between 18-23 MeV the $\alpha$-decay width is $90\%$ of
the total decay width and between 23-27 MeV $70\%$.  This reaction
mechanism is included neither in RPA nor in TBA.  This is probably the
reason why theory overestimates the peak height of the cross section
and does not reproduce the very broad experimental distribution.
 While the theoretical GQR cross section in $^{16}$O shows a well concentrated resonance, the theoretical monopole distribution is
very broad as no narrow single-particle resonances can contribute. It
resembles more the experimental pattern but is at least a factor of two too high in the resonance region. The situation is completely different for the GDR.  Our continuum
calculation reproduces nearly quantitatively the shape and magnitude of
the experimental distribution.  The reason is that the GDR is
dominated by the 1$\hbar\omega$ transitions which practically exhaust
the TRK-sum rule. However, the peak of the distribution are
  typically 1 MeV too low for the present Skyrme parameterization.

Figure \ref{fig:40Ca_dtba_ctba_m64k6} compares DTBA and CTBA
for the case of $^{40}$Ca. The agreement between theory and experiment
is very good for the GQR and GDR. In the case of the GMR our
theoretical distribution is about 2 MeV too high compared with the
experimental distribution.

\subsection{The dependence on the number of phonons}

In all the TBA calculations we use a large single-particle (s.p.)
basis both in the phonons and in the complex ($1p1h\otimes$phonon)
configurations, that is, a large number of $1p1h$ states in these
configurations.  As it was mentioned in Sec.~\ref{sec:calc}, the upper
limit for s.p. energies in all calculations for all nuclei was
$\varepsilon_{\text{max}}$ = 100 MeV. At the same time, only
collective phonons were used in the complex configurations.

\begin{figure*}
\centerline{\includegraphics[width=0.7\linewidth]
{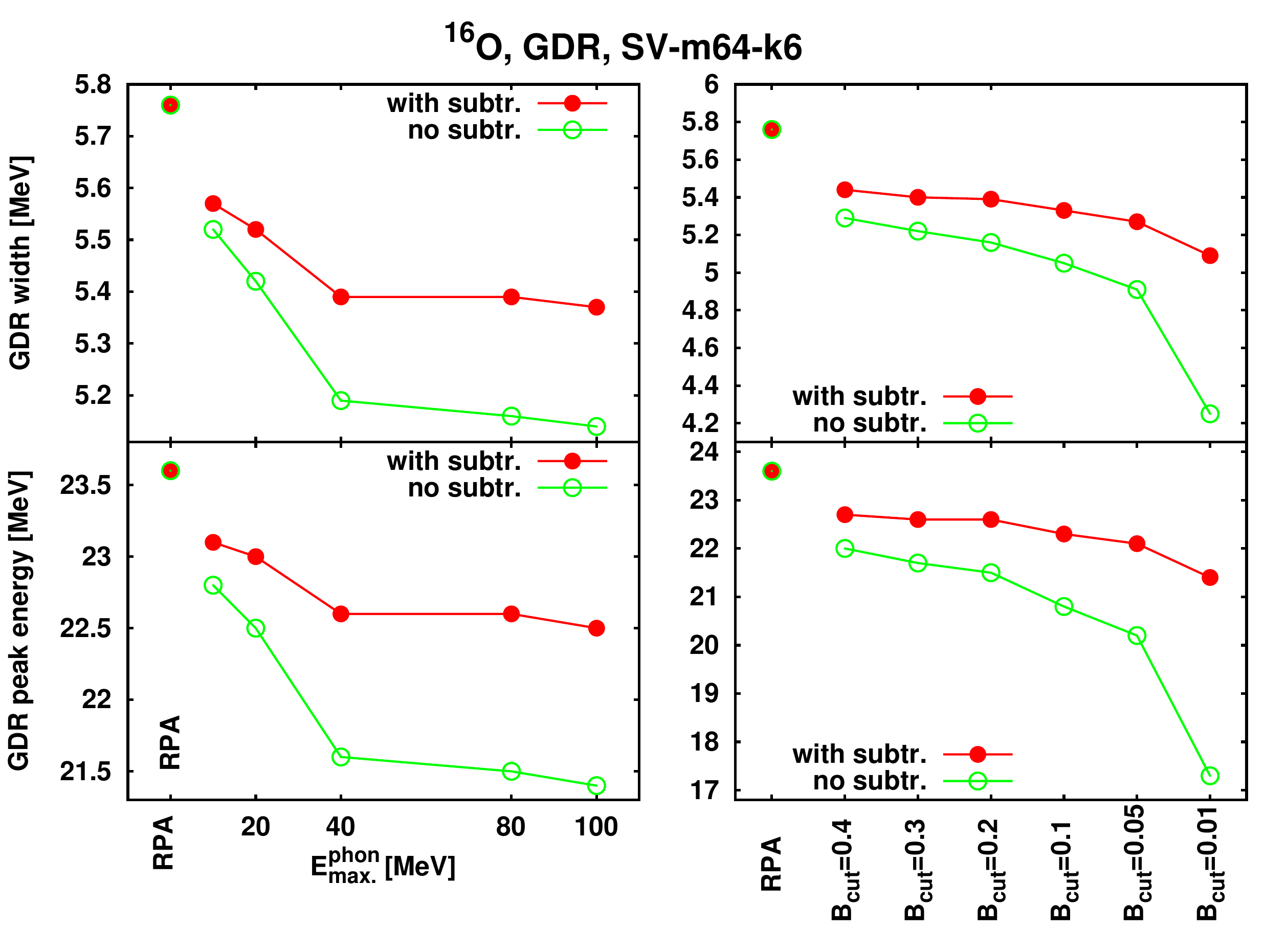}
}
\caption{\label{fig:dependences-16O-GDR}
Energy (upper part) and width (lower part) of the GDR in $^{16}$O
obtained from TBA calculations. The energy $E_0$ and the width $\Gamma$ are
the corresponding parameters of a Lorentzian fit to the theoretical results.
In the left corner of each figure the RPA result is given.
In the left column we present $E_0$ and $\Gamma$ as a function
of the maximal phonon energy $E^{\text{phon}}_{\text{max}}$ which we consider
in each calculation.
Table \ref{Nphonon_effect} gives the relation between energies
and the number of phonons for $B_{cut}$=0.2.
In the right column  we present $E_0$ and $\Gamma$ as function
of the minimal collectivity $B_{\text{cut}}$ of the phonons.
The maximal phonon energy is in all cases
$E^{\text{phon}}_{\text{max}}$ = 80 MeV.}
\end{figure*}

The dependence of the theoretical results on the number of phonons
used in the calculation is of crucial importance.  For this reason we
investigate this question in some detail.  The result of our
investigations for the GDR in $^{16}$O is summarized in
Fig. \ref{fig:dependences-16O-GDR}.  The energies $E_0$ and the widths
$\Gamma$ where derived from the theoretical cross section by a
Lorentzian fit.  We performed TBA calculations with and without the
subtraction procedure.  The two approaches give very different
results.  For comparison the RPA results are shown in the left upper
corner of each figure.

\begin{figure*}
\centerline{\includegraphics[width=0.7\linewidth]
{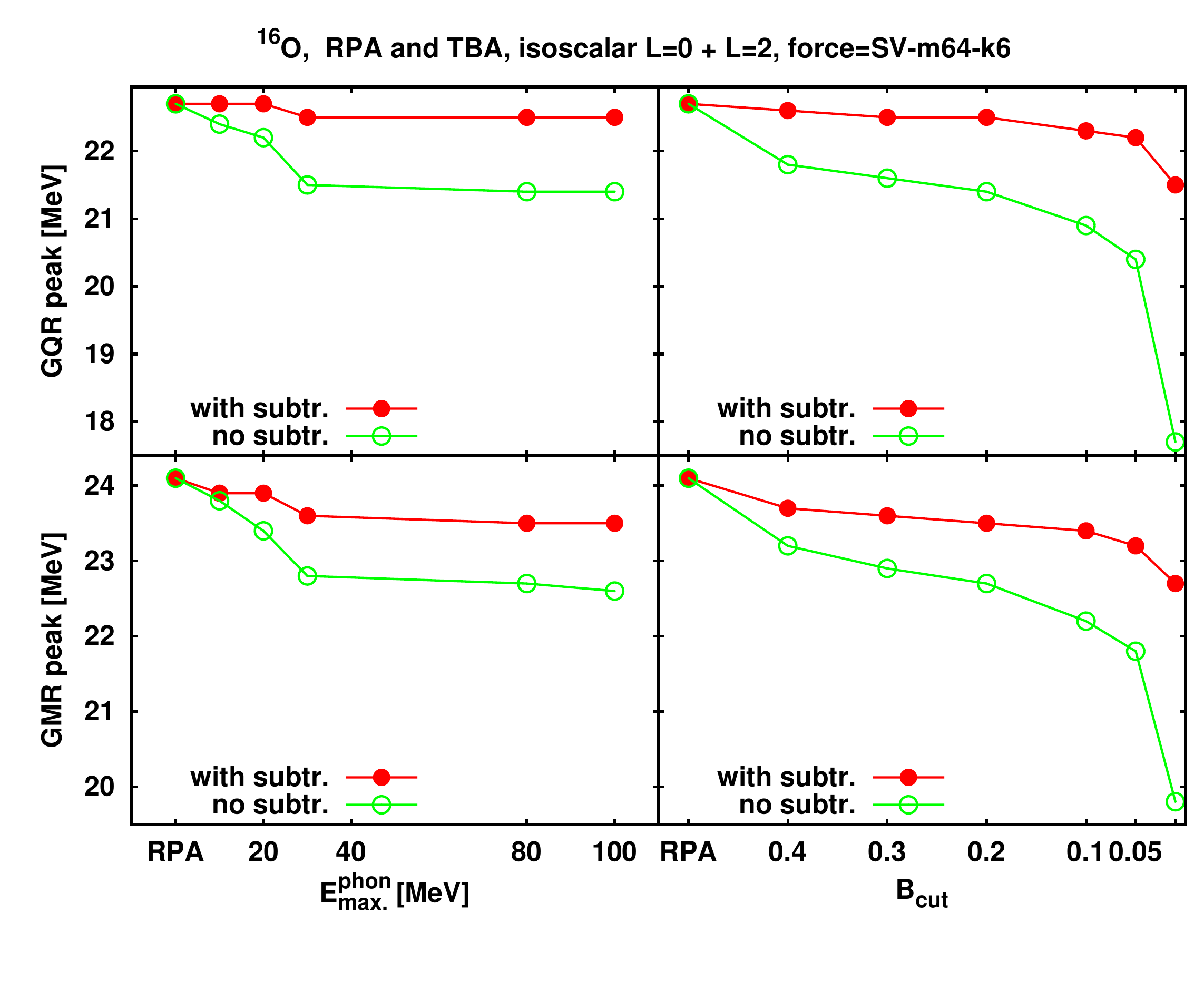}
}
\caption{\label{fig:dependences-MQ}
Same as in Fig.~\ref{fig:dependences-16O-GDR} but for the GMR and GQR in$^{16}$O.}
\end{figure*}

In the left column of Fig.~\ref{fig:dependences-16O-GDR}, the
dependence of $E_0$ and $\Gamma$ is presented as a function of the
maximal phonon energies $E^{\text{phon}}_{\text{max}}$.  From Table
\ref{Nphonon_effect}, one obtains the connection between
$E^{\text{phon}}_{\text{max}}$ and the number of phonons considered in
each calculation.  The single-particle basis in which we solve the RPA
and TBA equations includes s.p. states up to
$\varepsilon_{\text{max}}$ = 100 MeV and phonons up to the maximal
phonon energy $E^{\text{phon}}_{\text{max}}$ = 80 MeV.  In the right
column the same quantities are shown as a function of the lower cutoff
for transition strength $B_{\text{cut}}$ of the phonons where
\be
B_{\text{cut}} = B(EL)/B(EL)_{\text{max}} \,,
\ee
$B(EL)_{\text{max}}$ is the maximal reduced probability of the
excitation of the phonon states with the given angular momentum $L$.
The connection between $B_{\text{cut}}$ and the number of phonons can
be found again in Table \ref{Nphonon_effect}.  A too large number of
phonons causes two problems: violation of the Pauli principle and
double counting.  We reduce these problems as we restrict ourselves in
the actual calculations on phonons with  $B_{\text{cut}} \ge 0.2$.
Between $E^{\text{phon}}_{\text{max}}$ = 40 MeV and
$E^{\text{phon}}_{\text{max}}$ = 80 MeV the energy and width remain
stable if one applies the subtraction procedure.  This corresponds 55
phonons and 66 phonons, respectively (see Table~\ref{Nphonon_effect}
and the text).  In the right column the effect of an even larger
number of phonons is presented. Here the transition strength parameter
$B_{\text{cut}}$ ranges from 0.4 down 0.01. Here one sees strong
changes only for the extreme cases of $B_{\text{cut}}$ = 0.05 and
0.01.  The same is true also for the isoscalar resonances GMR and GQR
as can be seen in Fig.~\ref{fig:dependences-MQ}.  From this
investigation we conclude that our results in $^{16}$O are stable for
$\varepsilon_{\text{max}}$ = 100 MeV and 55 phonons.

\begin{table*}
\caption{\label{Nphonon_effect}
Relation between $E^{\text{phon}}_{\text{max}}$ and the number of phonons used in
$1p1h\otimes$phonon configurations for $^{16}$O.
The phonons were obtained in the $sp$ basis with
$\varepsilon_{\text{max}}$ = 100 MeV and angular momenta up to $L_{\text{max}} = 17$.
Only collective phonons were used in our actual TBA calculations, that is,
phonons with $B_{\text{cut}} \equiv B(EL)/B(EL)_{\text{max}}$ = 0.2
(see also the text).
Under these conditions, the number of phonons is fixed by the maximum
phonon energy $E^{\text{phon}}_{\text{max}}$.
The effect of the noncollective phonons is demonstrated for small
values $B_{\text{cut}}$ in the last two columns.}

\begin{ruledtabular}
\begin{tabular}{ccccccccccc}
$B_{\text{cut}}$              & 0.4 & 0.3 & 0.2 & 0.2 & 0.2 & 0.2 & 0.2 & 0.1 & 0.05 & 0.01 \\
$E^{\text{phon}}_{\text{max}}$& 80  & 80  & 10  & 20  & 40  & 80  & 100 & 80  & 80   & 80   \\
$N_{\text{phon}}$             & 42  & 52  & 1   &  6  & 55  & 66  & 69  & 117 & 166  & 325  \\
\end{tabular}
\end{ruledtabular}
\end{table*}

\begin{table}
\caption{\label{Emax_effect_208Pb}
Dependence of the resonance energy and width (Lorentzian parameters
$E_0$ and $\Gamma$) on the size of the sp basis used in
phonons and in $1p1h\otimes$phonon configurations for $^{208}$Pb with $B_{\text{cut}}$ = 0.2.
The size of the basis is characterized by the maximum energy
$\varepsilon_{\text{max}}$.For the GDR, the parameters were calculated photoabsorption cross section while for GMR and GQR the fractions EWSR were used. All the values are given in MeV.}

\begin{ruledtabular}
\begin{tabular}{cccccccc}
$\varepsilon_{\text{max}}$
               &      &\multicolumn{2}{c}{50}
                                    & \multicolumn{2}{c}{100}
                                                  & \multicolumn{2}{c}{150} \\
$E^{\text{phon}}_{\text{max}}$
               & RPA  &\multicolumn{2}{c}{40}
                                    & \multicolumn{2}{c}{40}
                                                  & \multicolumn{2}{c}{40} \\
subtract.      &      &  no  & yes  & no   & yes  & no   & yes  \\
\hline
               &      &      &      &      &      &      &      \\
GDR $E_0$      & 15.0 & 13.5 & 14.4 & 13.3 & 14.3 & 13.3 & 14.3 \\
$\Gamma$       & 4.60 & 4.63 & 4.57 & 4.61 & 4.53 & 4.63 & 4.54 \\
               &      &      &      &      &      &      &      \\
GMR $E_0$      & 14.4 & 13.3 & 14.1 & 13.1 & 14.0 & 13.0 & 13.9 \\
$\Gamma$       & 1.53 & 2.09 & 2.15 & 2.08 & 2.18 & 2.04 & 2.14 \\
               &      &      &      &      &      &      &      \\
GQR $E_0$      & 12.8 & 11.1 & 11.9 & 10.9 & 11.8 & 10.8 & 11.7 \\
$\Gamma$       & 1.04 & 1.10 & 1.13 & 1.10 & 1.19 & 1.10 & 1.24 \\
               &      &      &      &      &      &      &      \\
\end{tabular}
\end{ruledtabular}
\end{table}
In Table \ref{Emax_effect_208Pb} we compare again TBA results obtained
with and without the subtraction procedure as a function of the s.p.
space. Here we used $B_{\text{cut}}$ = 0.2 which corresponds
to 40 phonons. The results where the subtraction method was applied
are very stable.

\section{Results}
\label{sec:results}

From the huge variety of possible results, we concentrate on the three
most important giant resonances: the isoscalar giant monopole
resonance (GMR), the isoscalar giant quadrupole resonance (GQR), and
the isovector giant dipole resonance (GDR). For each resonances, we
consider mainly one number, the energy centroid taken in an energy
interval around the resonance peak. This serves as representative of
the peak energy. The energy centroids are computed as the ratio
$m_1/m_0$ (first versus zeroth energy moment of the corresponding
strengths). The moments are collected in exactly the same energy
windows which were used in the experimental averages.  We define a resonance peak energy by averaging the strength in a window
 around the resonance. The peak energy was defined as the energy
    centroid $m_1/m_0$ where the moments $m_1$ and $m_0$ were taken in a
    certain energy interval around the resonance peak. These windows are
    $11<E<40$ MeV for GMR and GQR in $^{16}$O, $15<E<30$ MeV for the
    GDR in $^{16}$O, $10 < E < 30$ MeV for GMR in $^{40, 48}$Ca,
    and $10 < E < 25$ MeV for GQR in $^{40, 48}$Ca, The centroids $E_0$ for the GDR
    in $^{40, 48}$Ca and for the GDR, GMR, and GQR in $^{208}$Pb
    were calculated in the window $E_0 \pm 2\delta $ where $\delta$ is the spectral
   dispersion (although with constraint $\delta \ge 2$ MeV).

\subsection{The impact of phonon coupling}

\begin{figure}
\centerline{\includegraphics[width=0.99\linewidth]{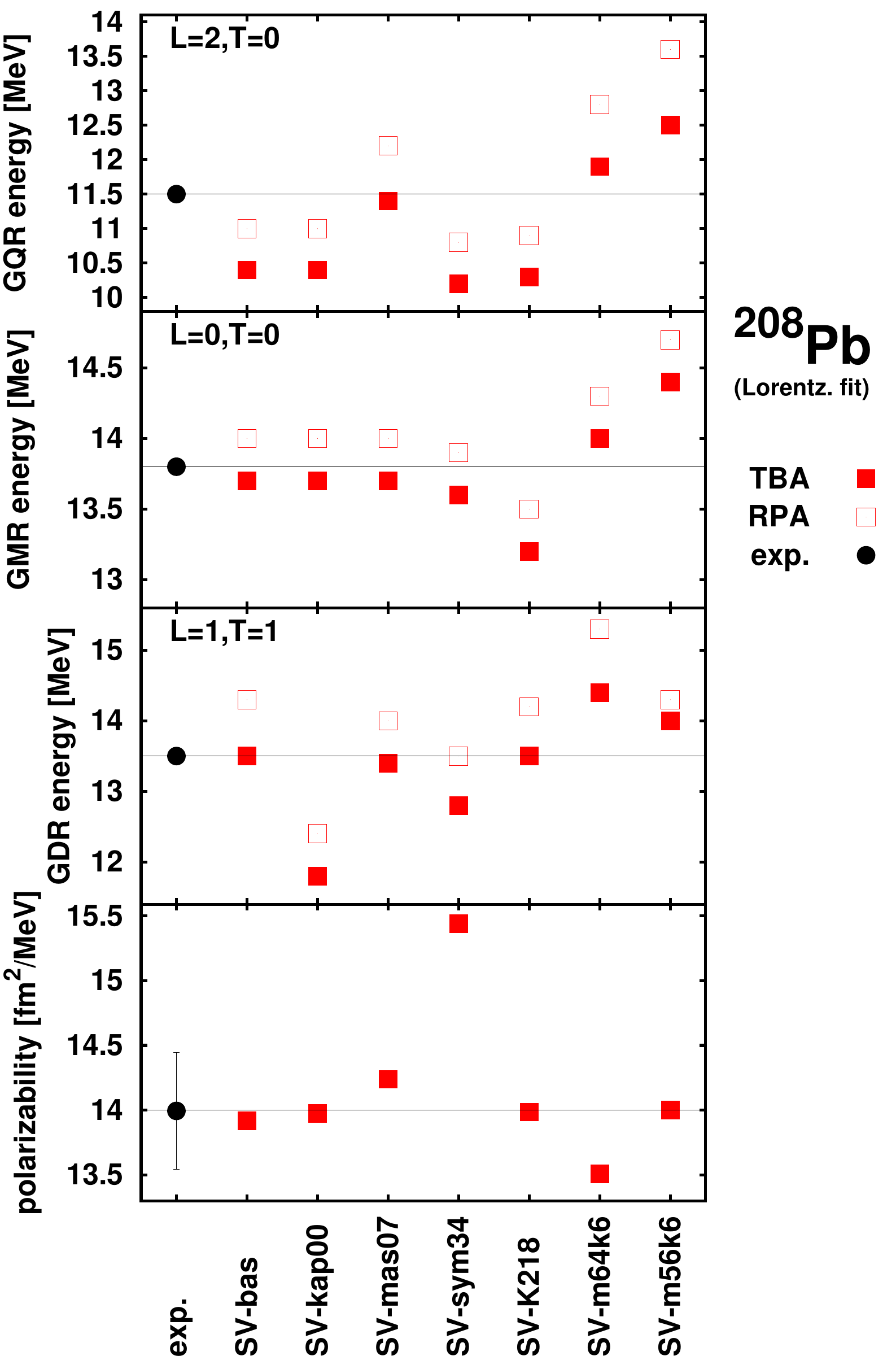}}
\caption{\label{fig:compare-208Pb}Comparison of giant resonance
  energies in $^{208}$Pb for a variety of Skyrme parameter sets as
  indicated. The energy centroids $E_0 = m_1/m_0$ are computed in the window $E_0 \pm 2\delta $ where $\delta$ is a dispersion (with the condition $\delta \ge 2$MeV).Open and filled symbols show the values calculated in the framework of RPA and TBA, respectively. The experimental data are taken from Refs.~\cite{Belyaev_1995} for the GDR, 
\cite{Youngblood_2004} for the GMR and the GQR, and \cite{Tamii_2011} for $\alpha_D$.}
\end{figure}
Fig. \ref{fig:compare-208Pb} summarizes the centroids for the three
major giant resonances in $^{208}$Pb (upper and middle) and the dipole
polarizability $\alpha_D$ (lower panel).  Let us briefly recall the
trends for RPA.  Changing $\kappa_\mathrm{TRK}$ affects almost
exclusively the GDR such that lower $\kappa_\mathrm{TRK}$ yields a
lower peak position.  Changing $m^*/m$ affects the GQR where lower
$m^*/m$ means higher peak position.  Changing $a_\mathrm{sym}$ affects
the dipole polarizability $\alpha_D$ with larger $a_\mathrm{sym}$
enhancing $\alpha_D$ although we see also a small side effect on
$\alpha_D$ from changing $m^*/m$.  Changing $K$ has an impact
predominantly on the GMR where lower $K$ lowers the peak energy.  The
combined changes of NMP in the two parametrizations SV-m64k6 and
SV-m56k6 yield changes in every mode.

The effect of the phonon coupling (move from open to closed symbols)
does not change these trends in general. The effects in details depend
very much on the actual parametrization but in all cases the energies
are shifted downwards. The lower panel of Fig. \ref{fig:compare-208Pb} shows the dipole polarizability $\alpha_D$. At
first glance, one misses the open symbols. The point is that the
polarizability represents a static response and TBA by virtue of the
subtraction method is designed such that it leaves stationary states
unchanged. Thus RPA and TBA results for $\alpha_D$ are exactly the
same which simplifies discussions in this case.  The large deviation
of $\alpha_D$ for SV-sym34 is the obvious effect of
$a_\mathrm{sym}$. It is noteworthy that the combination of changes on
NMP in SV-m56k6 cooperates to a good description of $\alpha_D$. Here,
the low $a_\mathrm{sym}$ alone would have produced a to low
$\alpha_D$. But the low $m^*/m$ drives $\alpha_D$ back up again.

\begin{figure}
\centerline{\includegraphics[angle=0,width=0.99\linewidth]{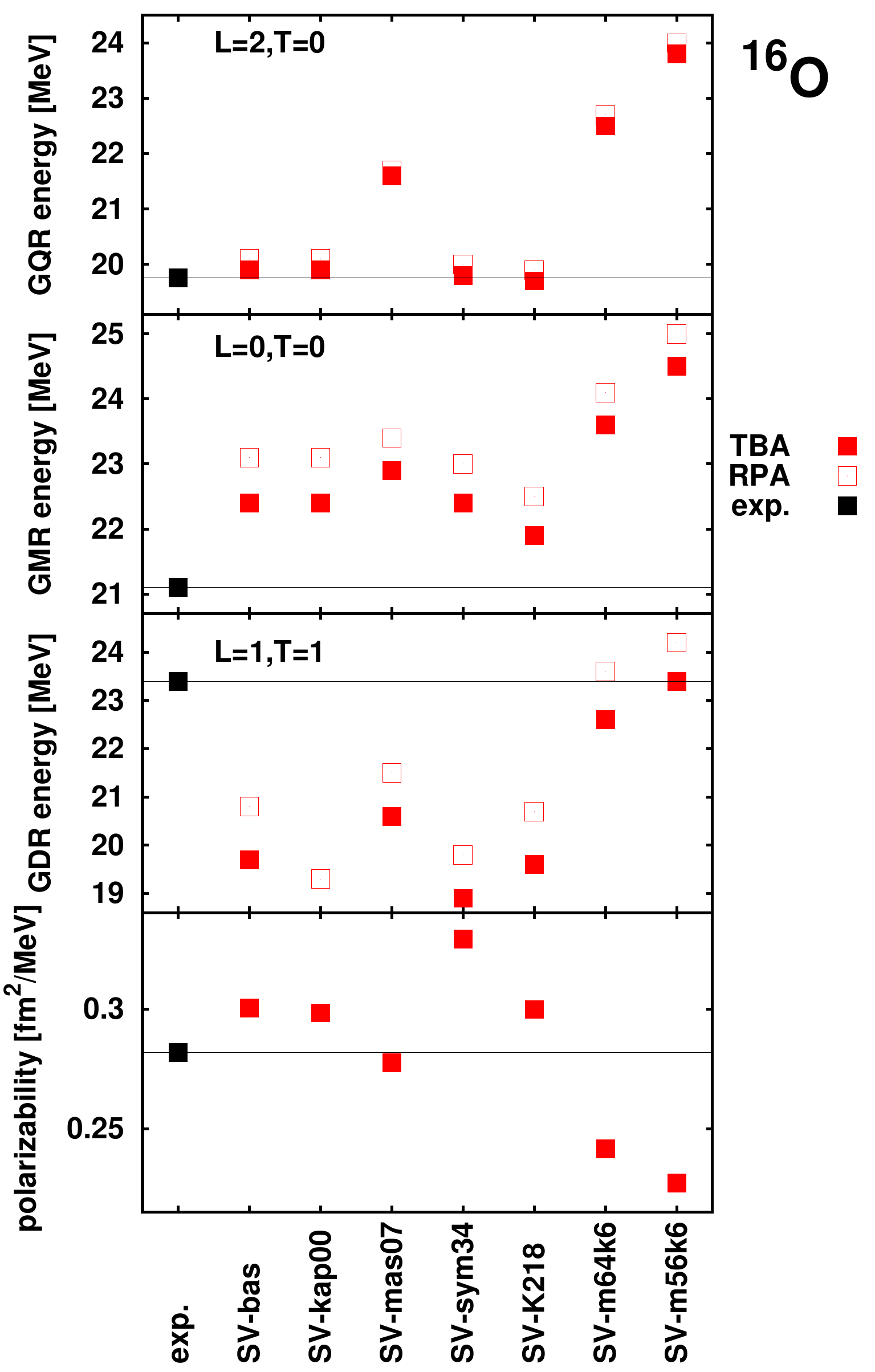}}
\caption{\label{fig:compare-16O-QTBA} 
As figure \ref{fig:compare-208Pb}, but for $^{16}$O.}
\end{figure}

Fig. \ref{fig:compare-16O-QTBA} shows the same for the light nucleus
$^{16}$O. As it is well known the standard Skyrme forces produce all
too low GDR energies (second panel from below) while those with
exotically low effective mass (SV-m56k6 and SV-m64k6) perform
fine. The situation is exactly opposite for the GQR (upper panel).
Here the standard forces do well and the exotic ones fail.  The GMR is
badly reproduced. All forces yield a too high centroid energy. As the
GMR and GQR are nearly continuously distributed the definition of a
centroid energy and a width is probably meaningless as it does not at
all characterize the experimental situation. To summarize the
situation one may conclude: For $^{208}$Pb alone, the conventional RPA
using the SV-bas parametrization manages to provide a good description
for all four features. However, SV-bas fails badly for the GDR in
$^{16}$O and to some extend also for the polarizability (the mismatch
of GMR is ignored here). It is only the new force SV-m56k6 in
combination with TBA which manages to get the GDR correct in both
nuclei \cite{Lyutorovich_2012}. But this spoils GMR, GQR, and
$\alpha_D(^{16}\mathrm{O})$. Considering the whole synopsis, we
realize that there is no force which reproduces all three giant
resonances and the polarizability simultaneously in $^{16}$O and
$^{208}$Pb, neither for RPA nor for TBA. Harmonizing all results
remains a challenge for future research. The situation in $^{40}$Ca
and $^{48}$Ca resembles more $^{208}$Pb as we have already seen in the
previous section. Therefore we may characterize the resonances by
centroid energy and a width.

 \begin{figure}
\centerline{\includegraphics[width=0.99\linewidth]{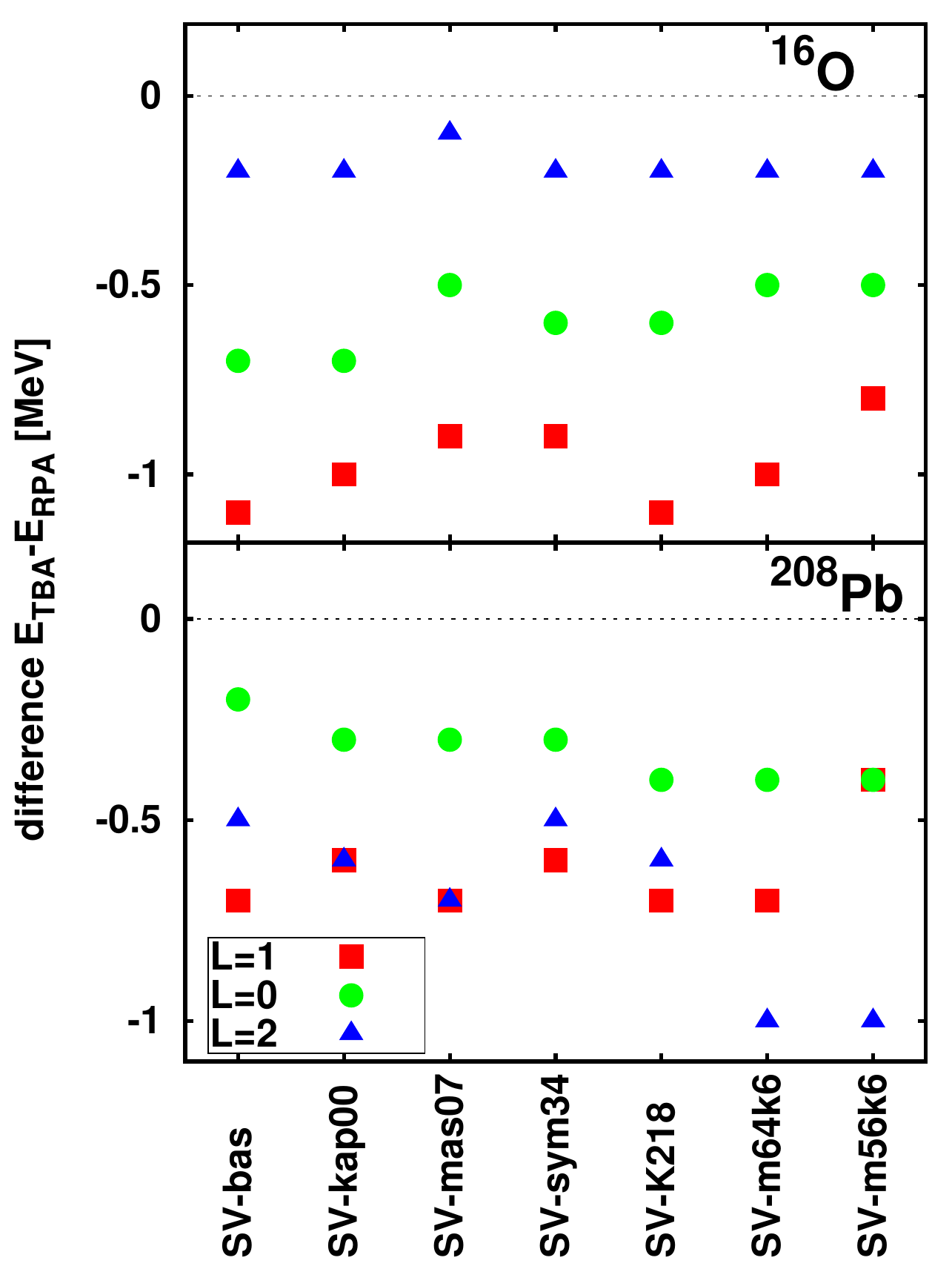}}
\caption{\label{fig:diff-208Pb} Difference between TBA and RPA
  for the giant resonance energies in $^{208}$Pb and $^{16}$O
  for a variety of Skyrme parametrizations as indicated.}
\end{figure}
Fig. \ref{fig:diff-208Pb} shows the differences of the energy
centroids between TBA and RPA for $^{208}$Pb and $^{16}$O. In this
figure the effects are much better presented than in the previous ones
where we showed the absolute values. in all cases the TBA energies are
lower than the RPA results. This is probably due to the first order
correction in the energy dependence of the effective mass discussed in
section II.A. The shifts are between one MeV for the GDR in $^{16}$O
and about 200 keV for the GQR in the same nucleus. The energy shift of
individual modes are always of the same magnitude.

\subsection{Final results compared with experiments}

The RPA and TBA theories work best in heavy nuclei where we have a
large s.p. basis which gives rise to very many low-lying and high-lying
collective phonons. This is the reason why in $^{208}$Pb for all
Skyrme parametrization we used, theory and experiment for all three
giant resonance modes are nearly in quantitative agreement as far as
the height of the cross sections and the widths are concerned.  We
recognize a strong reduction and the corresponding broadening of the
RPA cross sections due to the phonon coupling.  The mean energies of
the resonances on the other hand depend to some extend on the specific
Skyrme parametrization used.  This is also true for $^{40}$Ca and
$^{48}$Ca whereas in $^{16}$O only the GDR is well reproduced but not
the two isoscalar modes as we have already seen in the previous
chapter.

If we compare the shell structure of light nuclei such as $^{16}$O
with that of heavy mass nuclei such as $^{208}$Pb then one recognizes
that light nuclei posses only a very limited number of occupied states
which can support $1p1h$ excitations and thus a low density of
  $1p1h$ states.  In $^{208}$Pb, on the other hand, one has 126
  occupied neutron states and 82 proton states which all give rise to
  $1p1h$ excitations.  This leads to a high density of $1p1h$ states
  and subsequently rather smooth strength distributions already at the
  level of RPA. Moreover, light nuclei such as $^{12}$C and $^{16}$O
  contain a non-negligible amount of more complicated sub-structures
  as, e. g., $\alpha$-clusters. This is probably the reason, as
already discussed above and in Chapter III, that we can not reproduce the
  isoscalar modes.

\begin{figure*}
\centerline{\includegraphics[width=\linewidth]{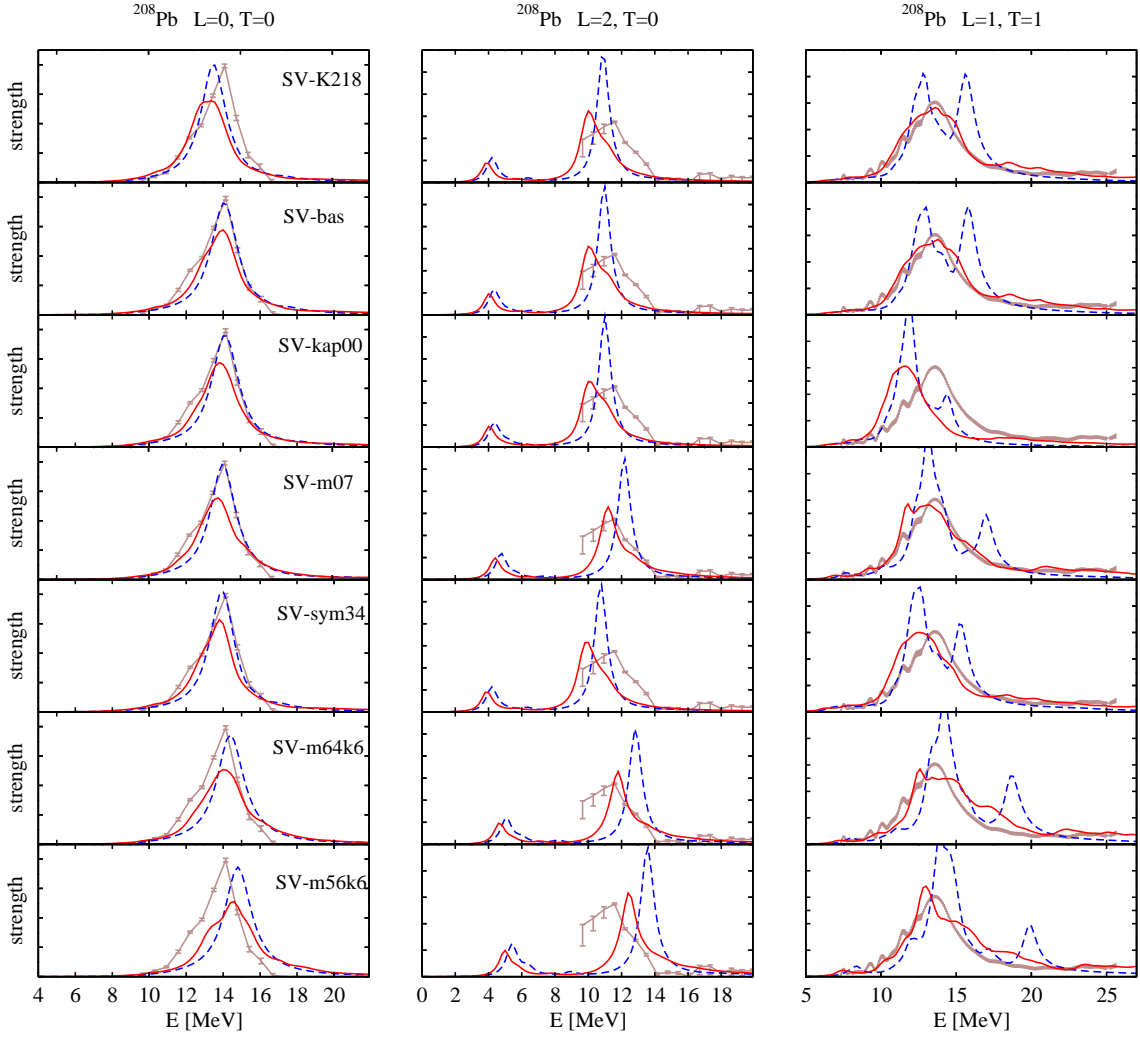}}
\caption{\label{fig:strength-all-208Pb} Detailed spectral strength
  distributions for $^{208}$Pb and the the three modes under consideration: 
  isoscalar monopole (left panels), isoscalar quadrupole (middle panels), 
  and isovector dipole (right panels).  
  Photo-absorption strength is shown in case of the dipole mode. 
  Results  are obtained with the seven Skyrme parametrizations 
  which we discussed in Chapter IIB. 
  Compared are strengths derived from RPA (blue dashed) and TBA 
  (full red) with experimental data (full brown) from \cite{Belyaev_1995} 
  for the GDR and \cite{Youngblood_2004} for the GMR and the GQR.}
\end{figure*}

In Fig.~\ref{fig:strength-all-208Pb}, the theoretical cross section of
GMR, GQR and GDR are compared with the experimental ones for
$^{208}$Pb.  The theoretical results are calculated with all seven
Skyrme parameter sets which we presented in Table~\ref{tab:NMP} of
Sec. II.B.  We first discuss the GMR which is closely connected with
the incompressibility $K$.  The first four parameter sets have an
incompressibility of 234 MeV.  The shape of the theoretical cross
sections and mean energies of all four parameter sets agree well with
the data except the peak height of the theoretical cross section is
slightly too low.  As three of the parameter set have the same
effective mass of $m^*/m=0.9$, it is not surprising that the
theoretical results are the same.  But also the fourth set (SV-mas07)
which has an effective mass of $m^*/m=0.7$ yields essentially the same
cross section.  The largest difference delivers set SV-m56k6 with an
effective mass of $m^*/m=0.56$.  Here the theoretical peak in the
cross section is about 1.5 MeV to high.

\begin{figure*}
\centerline{\includegraphics[width=\linewidth]{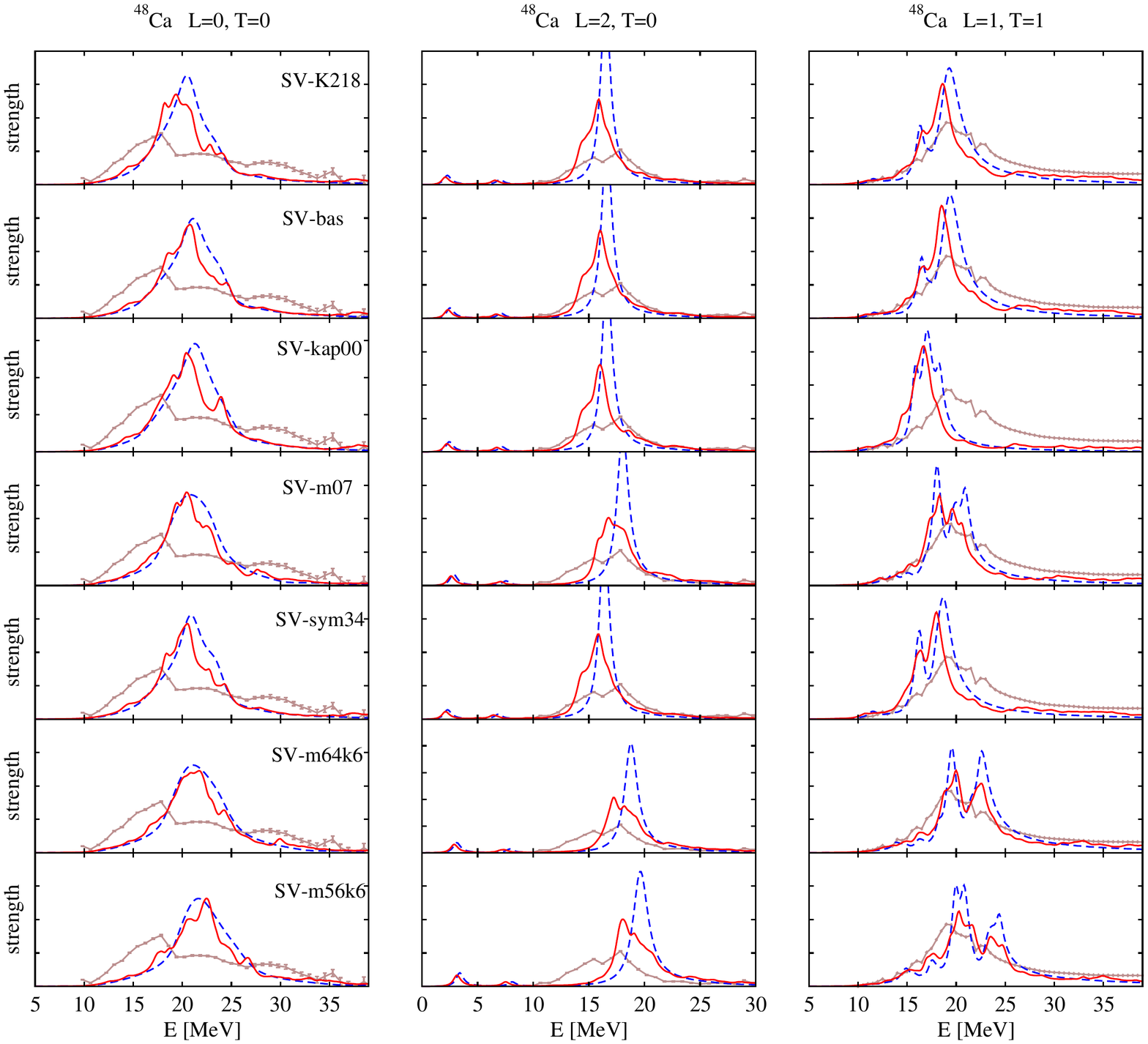}}
\caption{\label{fig:strength-all-48Ca} Same results as in the previous figure but for
$^{48}$Ca. The data are taken from \cite{Erokhova_2003} for the GDR and from \cite{Anders_2013} for the GMR and the GQR.}
\end{figure*}

In Fig.~\ref{fig:strength-all-48Ca} we compare our theoretical results
for $^{48}$Ca with the data. The GDR with the specifically adjusted
parameter sets \cite{Lyutorovich_2012} to reproduce the GDR in
$^{208}$Pb and $^{16}$O shown in the last two rows agree nicely with
the data. For the other parameter sets the agreement is also not
bad. The height of the theoretical cross section for all isoscalar
resonances are roughly a factor two too large compared with the
experimental ones. Here we have to bear in mind that also deep-lying
hole states are important which are very broad. Their widths are
insufficiently described by RPA phonons alone and therefore the
theoretical resonances are too narrow.

\section{Summary}

The present paper is an extended version of a previous short note
  \cite{Lyutorovich_2015}.  It is concerned with the time-blocking
  approximation (TBA) which is an extension of the widely used
  random-phase approximation (RPA) by complex configurations in terms
  of $1p1h$ states coupled to RPA phonons and addresses a couple of
  basic questions in this scheme: proper treatment of the continuum,
  restoration of stability of ground and excited states, and size of
  phonon space.

First, we explain here details of the self-consistent continuum TBA
which is a new method for handling the single-particle
continuum. This method had been further developed to include also the
spin-orbit contribution such that our new calculations are fully
self-consistent. We then present numerical results which demonstrate
the advantages of the continuum treatment as compared to the
conventional treatment in a discrete basis.

The phonon coupling modifies the residual two-body interaction which,
in principle, would require to compute a new, correlated ground state
in order to stay consistent and to achieve a stable excitation
spectrum with non-imaginary excitation energies. However, this would
introduce a double counting because most ground-state correlations
are already incorporated in an effective mean-field theory. The
problem is solved by the subtraction scheme, subtracting the stationary
(zero-frequency) part of the effective interaction. This leaves the
ground state unchanged and delivers stable excitations throughout.
It also helps to achieve convergence with phonon number.

A long standing problem concerns the stability of the TBA with respect
to the choice of the number of phonons and the size of the single
particle space. Here we present the results of detailed calculations
with systematically scanned numbers of phonons. An important result is
that the energies and widths are stable over a large range if the
subtraction method is included in the TBA. This identifies a window
of phonon numbers where the results are robust.

Having a well tested numerical scheme for (continuum) RPA and TBA at
hand, we investigate the dependence of the three main giant resonances
on the basic properties of a Skyrme parameterization:
incompressibility, iso-scalar effective mass, symmetry energy and TRK
sum rule enhancement.  And we do that for RPA in comparison to TBA.
TBA generally down-shifts the peak resonance energies by up to 1
MeV. The shift is about same for all parameterizations for a given mode
and nucleus. It differs for the three modes and also depends on the
nucleus. Although  the results show a reasonable general agreement with the 
data, a parameterization
which is able to describe equally well all three resonance modes in
heavy as well as light nuclei has not been found.

\begin{acknowledgments}
This work has been supported by contract Re322-13/1 from the DFG. N.L. and V.T. acknowledge 
St.Petersburg State University for a research grand 11.38.648.2013. N.L. acknowledges 
St.Petersburg State University for a research grand 11.38.193.2014. We thank S.P. Kamerdzhiev for fruitful discussions and Dave Youngblood for providing us with experimental data. Research was supported by Resource Center "Computer Center" of SPbU.
\end{acknowledgments}

\appendix

\section{Continuum in a discrete basis representation}
\label{app:cont}

In the RPA and TBA the response function $R(\omega)$ is a solution of the
Bethe-Salpeter equations (\ref{rfrpa}) and (\ref{rftba}), respectively.
The propagator $R^{(0)}_{\vphd}(\omega)$ in these equations
in the discrete basis representation has the form
\be
R^{(0)}_{ph,p'h'}(\omega) = - \frac{\delta^{\vphu}_{pp'}\delta^{\vphu}_{h'h}}
{\omega - \ve^{\vphu}_{ph}}\,,
\label{r0ph}
\ee
\be
R^{(0)}_{hp,h'p'}(\omega) = \frac{\delta^{\vphu}_{p'p}\delta^{\vphu}_{hh'}}
{\omega + \ve^{\vphu}_{ph}}\,,
\label{r0hp}
\ee
where $\ve^{\vphu}_{ph} = \ve^{\vphu}_{p} - \ve^{\vphu}_{h}$.

Let us represent Eqs. (\ref{r0ph}) and (\ref{r0hp}) in the form
\be
R^{(0)}_{ph,p'h'}(\omega) = - \delta^{\vphu}_{h'h}\,
\langle\,p\,|\,G^{\mbsu{MF}(+)}(\ve^{\vphu}_{h} + \omega)\,|\,p' \rangle,
\label{r0phg}
\ee
\be
R^{(0)}_{hp,h'p'}(\omega) = - \delta^{\vphu}_{hh'}\,
\langle\,p'|\,G^{\mbsu{MF}(+)}(\ve^{\vphu}_{h} - \omega)\,|\,p\,\rangle,
\label{r0hpg}
\ee
where
\be
G^{\mbsu{MF}(+)}(\ve) = G^{\mbsu{MF}}(\ve)
- \sum_{h}\frac{|\,h\,\rangle \langle\,h\,|}{\ve - \ve^{\vphu}_{h}}\,,
\label{defgpls}
\ee
$G^{\mbsu{MF}}(\ve)$ is the single-particle mean-field Green function,
$|\,p\,\rangle$ and $|\,h\,\rangle$ are the single-particle
wave functions of particles and holes.
The superscript $(+)$ in the notation $G^{\mbsu{MF}(+)}(\ve)$ means that
this function has the poles only above Fermi level.
The equivalence of Eqs. (\ref{r0ph})--(\ref{r0hp}) and
(\ref{r0phg})--(\ref{r0hpg}) follows from the spectral expansion
\be
G^{\mbsu{MF}}(\ve) =
\sum_{h}\frac{|\,h\,\rangle \langle\,h\,|}{\ve - \ve^{\vphu}_{h}} +
\sum_{p}\frac{|\,p\,\rangle \langle\,p\,|}{\ve - \ve^{\vphu}_{p}}
\label{spgspectr}
\ee
and the orthonormality of the wave functions of the discrete basis.

The discrete basis in this scheme
is defined as a complete set of solutions of the
Schr\"odinger equation with the box boundary conditions (b.b.c.).
Let us introduce another complete set of solutions of this equation
obtained by imposing continuum wave boundary conditions (c.b.c.).
This set includes a finite number of the discrete states of holes
and particles and a particle continuum.
Respective mean-field Green functions and the single-particle states
will be denoted as $\tilde{G}^{\mbsu{MF}}(\ve)$,
$|\,\tilde{p}\,\rangle$ and $|\,\tilde{h}\,\rangle$.

The method of inclusion of the continuum in the discrete basis
representation consists in the replacement of the uncorrelated
$ph$ propagator $R^{(0)}_{\vphd}(\omega)$ in Eqs. (\ref{rfrpa}) and (\ref{rftba})
by the propagator
$\tilde{R}^{(0)}_{\vphd}(\omega)$, which is defined by the formulas:
\be
\tilde{R}^{(0)}_{ph,p'h'}(\omega) = - \delta^{\vphu}_{h'h}\,
\langle\,p\,|\,\tilde{G}^{\mbsu{MF}(+)}(\ve^{\vphu}_{h} + \omega)\,
|\,p' \rangle,
\label{r0phgc}
\ee
\be
\tilde{R}^{(0)}_{hp,h'p'}(\omega) = - \delta^{\vphu}_{hh'}\,
\langle\,p'|\,\tilde{G}^{\mbsu{MF}(+)}(\ve^{\vphu}_{h} - \omega)\,
|\,p\,\rangle,
\label{r0hpgc}
\ee
\be
\tilde{G}^{\mbsu{MF}(+)}(\ve) = \tilde{G}^{\mbsu{MF}}(\ve) -
\sum_{\tilde{h}}\frac{|\,\tilde{h}\,\rangle \langle\,\tilde{h}\,|}
{\ve - \ve^{\vphu}_{\tilde{h}}}\,.
\label{defgplsc}
\ee
Eqs. (\ref{r0phgc})--(\ref{r0hpgc}) are obtained from Eqs.
(\ref{r0phg})--(\ref{r0hpg}) by the replacement of the function
$G^{\mbsu{MF}(+)}(\ve)$ by the function $\tilde{G}^{\mbsu{MF}(+)}(\ve)$.
The Green function $\tilde{G}^{\mbsu{MF}}(\ve)$ in Eq.~(\ref{defgplsc})
is calculated via the regular and irregular solutions of the
Schr\"odinger equation (with c.b.c.)
by means of the known technique \cite{Shlomo_1975}.
The matrix elements of $\tilde{G}^{\mbsu{MF}(+)}(\ve)$
are calculated with particle wave functions $|\,p\,\rangle$ and
$|\,p'\rangle$ of the discrete basis.
Thus, the RPA and the TBA equations (\ref{rfrpa}) and (\ref{rftba})
are solved in the discrete basis representation.
However, in contrast to the initial uncorrelated $ph$ propagator
$R^{(0)}_{\vphd}(\omega)$, the propagator
$\tilde{R}^{(0)}_{\vphd}(\omega)$
does not contain the {\it discrete} poles $\omega = \pm\ve^{\vphu}_{ph}$
corresponding to the transitions between the hole states and the
{\it discrete} particle states with positive energies,
since these states are replaced by the continuum included in
the Green function $\tilde{G}^{\mbsu{MF}}(\ve)$.

This method recovers the exact method \cite{Shlomo_1975} of treatment
of the continuum in the coordinate representation if the discrete
basis is sufficiently complete and the radius of the box is
sufficiently large to ensure the equality
$|\,h\,\rangle = |\,\tilde{h}\,\rangle$.

As a criterion of the fulfillment of this equality we choose
the absolute value of the difference between the energies of the hole states calculated
with continuum wave boundary and box boundary conditions, respectively:
$\Delta \ve^{\vphu}_h = \ve^{\vphu}_h - \ve_{\tilde{h}}$.
In all our calculations (with $R_{\mbsu{box}}=15$~fm for
$^{16}$O, $^{40}$Ca, and $^{48}$Ca and $R_{\mbsu{box}}=18$~fm for $^{132}$Sn and 
$^{208}$Pb) we have $\max |\Delta \ve^{\vphu}_h| \lesssim 10^{-5}$ MeV.
\

\bibliographystyle{apsrev4-1}
\bibliography{A-GMQR-5}

\end{document}